\documentclass[useAMS,usenatbib,usegraphicx]{mn2e}

\usepackage{bm}

\newcommand{\dd}{\mathrm{d}} 
\newcommand{\bra}[1]{\langle #1|}
\newcommand{\ket}[1]{|#1\rangle}
\newcommand{\kB}{k_\mathrm{B}}

\newcommand{\be}{\begin{equation}}
\newcommand{\ee}{\end{equation}}
\newcommand{\ba}{\begin{eqnarray}}
\newcommand{\ea}{\end{eqnarray}}
\newcommand{\omc}{\omega_\mathrm{c}}
\newcommand{\Omc}{\Omega_\mathrm{c}}
\newcommand{\req}[1]{Eq.~(\ref{#1})}

\title[Radiative transitions of the helium atom]
{Radiative transitions of the helium atom in highly magnetized
neutron star atmospheres}

\author[Z. Medin, D. Lai, and A. Y. Potekhin]{Z. Medin$^{1}$, D. Lai$^{1}$, and A. Y. Potekhin$^{1,2}$\\
$^{1}$Center for Radiophysics and Space Research, Department of Astronomy, Cornell University, Ithaca, NY 14853\\
$^{2}$Ioffe Physico-Technical Institute, Politekhnicheskaya 26, 194021 Saint-Petersburg, Russia}

\begin{document}

\date{Accepted 2007 September 26. Received 2007 May 21; in original form 2007 April 12}

\pagerange{\pageref{firstpage}--\pageref{lastpage}} \pubyear{2008}

\label{firstpage}

\maketitle

\begin{abstract}
Recent observations of thermally emitting isolated neutron stars
revealed spectral features that could be interpreted as radiative
transitions of He in a magnetized neutron star atmosphere.  We present
Hartree--Fock calculations of the polarization-dependent
photoionization cross sections of the He atom in strong magnetic fields
ranging from $10^{12}$~G to $10^{14}$~G. Convenient fitting formulae
for the cross sections are given as well as related oscillator
strengths for various bound-bound transitions.  The effects of finite
nucleus mass on the radiative
absorption cross sections are examined using perturbation theory.
\end{abstract}


\section{Introduction}
\label{sec:intro}

An important advance in neutron star astrophysics in the last few
years has been the detection and detailed studies of surface emission
from a large number of isolated neutron stars (NSs), including radio
pulsars, magnetars, and radio-quiet NSs
\cite[e.g.,][]{Kaspi06,Harding06}. This was made possible by X-ray
telescopes such as {\it Chandra} and {\it XMM-Newton}. Such studies
can potentially provide invaluable information on the physical
properties and evolution of NSs (e.g., equation of state at
super-nuclear densities, cooling history, surface magnetic field and
composition).
Of great interest are the radio-quiet, thermally emitting NSs
\citep[e.g.,][]{haberl06b}: they share the common property that their spectra
appear to be entirely thermal, indicating
that the emission arises directly from the NS surfaces,
uncontaminated by magnetospheric emission.
The true nature of these sources, however, is unclear at present:
they could be young cooling NSs, or NSs kept hot by accretion
from the ISM, or magnetar descendants.
While some of these NSs (e.g., RX J1856.5$-$3754) have featureless
X-ray
spectrum remarkably well described by blackbody
\citep[e.g.,][]{Burwitz03}
or by emission from a condensed surface covered by a thin atmosphere
\citep{hoetal07},
 a single or multiple absorption
features at $E\simeq 0.2$--$1$~keV have been detected from several
sources \citep[see][]{vankerkwijk07}: e.g., 1E 1207.4$-$5209
(0.7 and 1.4~keV, possibly also 2.1,~2.8~keV;
\citealt{sanwal02,deluca04,mch05}), RX J1308.6+2127 (0.2--0.3~keV;
\citealt{haberl03}), RX J1605.3+3249 (0.45~keV;
\citealt{vankerkwijk04}), RX J0720.4$-$3125 (0.27~keV;
\citealt{haberl06}),
and possibly RBS 1774 ($\sim0.7$~keV; \citealt{zane05}).
   The identifications
of these features, however, remain uncertain, with suggestions ranging
from proton cyclotron lines to atomic transitions of H, He, or mid-Z
atoms in a strong magnetic field
\citep[see][]{sanwal02,ho04,PB05,moriho07}. Clearly, understanding
these absorption lines is very important as it would lead to direct
measurement of the NS surface magnetic fields and compositions,
shedding light on the nature of these objects. Multiple lines also
have the potential of constraining the mass-radius relation of NSs
(through measurement of gravitational redshift).

Since the thermal radiation from a NS is mediated by its
atmosphere (if $T$ is sufficiently high so that the surface does
not condense into a solid; see, e.g.,
\citealt{vanadelsberg05,medin06,medin07}), detailed
modelling of radiative
transfer in magnetized NS atmospheres is important.
 The atmosphere
composition of the NS is unknown {\it a priori}.  Because of the
efficient gravitational separation of light and heavy elements, a
pure H atmosphere is expected even if a small amount of
fallback or accretion occurs after NS formation.  A pure He atmosphere
results if H is completely burnt out, and a heavy-element (e.g.,
Fe) atmosphere may be possible if no fallback/accretion occurs.
The atmosphere composition may also be affected by (slow)
diffusive nuclear burning 
in the outer NS envelope \citep*{chang04}, as well as by the
bombardment on the surface by fast particles from NS magnetospheres
\citep[e.g.,][]{beloborodov07}.
Fully ionized atmosphere models in various
magnetic field regimes have been extensively studied
\citep[e.g.,][]{shib92,zane01,ho01},
 including the effect of vacuum
polarization \citep[see][]{ho03,lai02,lai03,vanadelsberg06}.  Because
a strong magnetic field greatly increases the binding energies of
atoms, molecules, and other bound species
\citep[for a review, see][]{lai01}, these
bound states may have appreciable abundances in the NS atmosphere,
as guessed by
\citet*{CLR70} and confirmed by calculations of 
\citet{lai97} and \citet*{pcg99}.
Early considerations of partially ionized and strongly magnetized
atmospheres \citep*[e.g.,][]{RRM} relied on oversimplified treatments
of atomic physics
and plasma thermodynamics (ionization equilibrium,
equation of state, and nonideal plasma effects).
  Recently, a
thermodynamically consistent equation of state and opacities for
magnetized ($B=10^{12}-10^{15}$~G), partially ionized H plasma have
been obtained \citep{pc03,pc04}, and the effect of bound atoms on the
dielectric tensor of the plasma has also been studied
\citep{plch04}. These improvements have been incorporated into
partially ionized, magnetic NS atmosphere models
\citep{hlpc03,hoetal07,plch04,pclhv06}.
 Mid-$Z$ element atmospheres for $B\sim
10^{12}-10^{13}$~G were recently studied by \citet{moriho07}.

In this paper we focus on He atoms and their radiative transitions in
magnetic NS atmospheres. It is well known that for $B\gg Z^2 B_0$, where
$Z$ is the charge number of the nucleus and $B_0 = e^3 m_e^2/\hbar^3 c =
2.35\times10^9$~G, the binding energy of an atom is significantly
increased over its zero-field value.  In this \emph{strong-field regime}
the electrons are confined to the ground Landau level, and one may apply
the \emph{adiabatic approximation}, in which  electron motions along and
across the field are assumed to be decoupled from each other (see
Sect.\,\ref{sect:bound}).  Using this approximation in combination with
the Hartree--Fock method (``1DHF approximation''), several groups
calculated binding energies for the helium atom
\citep*{Proschel-ea,thurner-ea} and also for some other atoms and
molecules \citep*{neuhauser86,neuhauser87,miller-neu,lai92}.
\citet{MoriHailey} developed a ``multiconfigurational perturbative
hybrid Hartree--Fock'' approach, which is a perturbative improvement of
the 1DHF method.  Other methods of calculation include 
Thomas--Fermi-like models
\citep[e.g.,][]{AS91},
the density
functional theory \citep[e.g.,][]{RelovskyRuder,medin06}, variational
methods  \citep[e.g.,][]{muller84,vb89,joc99,turbiner}, and 2D
Hartree--Fock mesh calculations \citep{Ivanov94,IvanovSchm00}
which do not directly employ the adiabatic approximation.

In strong magnetic fields, the finite nuclear mass and centre-of-mass
motion affect the atomic structure in a nontrivial way (e.g.,
\citealt{lai01}; see Sect.\,\ref{sect:motion}).
The stronger $B$ is,
the more important the effects of finite nuclear mass are. Apart from
the H atom, these effects have been calculated only for the He atom
which \emph{rests as a whole}, but has a moving nucleus
\citep{hujaj1,hujaj2}, and for the He$^+$ ion \citep*{BPV,PB05}.

There were relatively few publications devoted to radiative
transitions of non-hydrogenic atoms in strong magnetic fields. Several
authors \citep{miller-neu,thurner-ea,joc99,MoriHailey,hujaj2}
calculated oscillator strengths for bound-bound transitions;
\citet{miller-neu} presented also a few integrated bound-free
oscillator strengths. \citet{RRM} calculated opacities of strongly
magnetized iron, using photoionization cross sections obtained by
M.~C. Miller (unpublished). To the best of our knowledge, there were
no published calculations of polarization-dependent photoionization
cross sections for the He atom in the strong-field regime, as well as
the calculations of the atomic motion effect on the photoabsorption
coefficients for He in this regime.
Moreover, the subtle effect of
exchange interaction involving free electrons and the possible role
of two-electron transitions (see Sect.\,\ref{sect:photo}) were not
discussed before.

In this paper we perform detailed calculations of radiative
transitions of the He atom using the 1DHF approximation.  The total
error introduced into our calculations by the use of these two
approximations, the Hartree-Fock method and the adiabatic
approximation, is of order $1\%$ or less, as can be seen by the
following considerations: The Hartree-Fock method is approximate
because electron correlations are neglected. Due to their mutual
repulsion, any pair of electrons tend to be more distant from each
other than the Hartree-Fock wave function would indicate. In
zero-field, this correlation effect is especially pronounced for the
spin-singlet states of electrons for which the spatial wave function
is symmetrical. In strong magnetic fields ($B\gg B_0$), the electron
spins (in the ground state) are all aligned antiparallel to the
magnetic field, and the multielectron spatial wave function is
antisymmetric with respect to the interchange of two electrons. Thus
the error in the Hartree-Fock approach is expected to be less than the
$1\%$ accuracy characteristic of zero-field Hartree-Fock calculations
(\citealt{neuhauser87}; \citealt*{schmelcher99}; for $B=0$ see
\citealt{scrinzi98}). The adiabatic approximation is also very
accurate at $B\gg Z^2 B_0$. Indeed, a comparision of the ground-state
energy values calculated here to those of \citet{Ivanov94} (who did
not use the adiabatic approximation) shows an agreement to within
$1\%$ for $B=10^{12}$~G and to within $0.1\%$ for $B=10^{13}$~G.

The paper is organized as follows. Section \ref{sect:stat} describes
our  calculations of the bound states and continuum states of the He atom,
and section \ref{sec:transitions} contains relevant equations for
radiative transitions. We present our numerical results and fitting
formulae in section \ref{sect:res} and examine the effects of finite
nucleus mass on the photoabsorption cross sections in section
\ref{sect:motion}.

\section{Bound states and singly-ionized states of helium atoms in
strong magnetic fields}
\label{sect:stat}

\subsection{Bound states of the helium atom}
\label{sect:bound}

To define the notation, we briefly describe 1DHF
calculations for He atoms in strong magnetic fields. Each electron in the
atom is described by a one-electron wave function (orbital). 
If the magnetic field is sufficiently strong
(e.g., $B\gg10^{10}$~G for He ground state),
the motion of an electron perpendicular to the magnetic field lines
is mainly governed by the Lorentz force, which is, on the average,
stronger than the Coulomb force.
In this case, the adiabatic approximation can be employed -- i.e.,
the wave function
can be separated into a transverse (perpendicular to the external
magnetic field) component and a longitudinal (along the magnetic
field) component:
\be
\phi_{m\nu}(\bm{r}) = f_{m\nu}(z) W_m(\bm{r}_\perp) \,.
\ee
Here $W_m$ is the ground-state Landau wave function \citep[e.g.,][]{landau77} given by
\be
W_m(\bm{r}_\perp) = \frac{1}{\rho_0 \sqrt{2\pi m!}}
\left(\frac{\rho}{\sqrt{2}\rho_0}\right)^m
\exp\left(\frac{-\rho^2}{4\rho_0^2}\right)
\mathrm{e}^{-\mathrm{i}m\varphi} ,
\label{Wmeq}
\ee
where 
$(\rho,\varphi)$ are the polar coordinates of $\bm{r}_\perp$,
$\rho_0=(\hbar c/eB)^{1/2}$ is the 
magnetic length
and
$f_{m\nu}$ is the longitudinal wave function which
can be calculated
numerically. The quantum number $m$ ($\ge 0$
for the considered ground Landau state) specifies 
the \emph{negative}
of the $z$-projection of the electron orbital angular momentum.
We restrict our consideration to electrons in the ground Landau level;
for these electrons, $m$ specifies also
the (transverse)
distance of the guiding centre of the electron from the ion,
$\rho_m=(2m+1)^{1/2} \rho_0$. The quantum number $\nu$ specifies the
number of nodes in the longitudinal wave function.
The spins of the electrons are taken to be
aligned anti-parallel with the magnetic field, and so do not enter
into any of our equations.
In addition, we assume that the ion is completely stationary (the
`infinite ion mass' approximation). In general, the latter assumption
is not necessary for the applicability of the adiabatic approximation
(see, e.g., \citealt{P94}). The accuracy of the infinite ion mass
approximation will be discussed in Sect.\,\ref{sect:motion}.

Note that we use non-relativistic quantum mechanics in our calculations, 
even when $\hbar\omega_{Be}\ga m_ec^2$ or 
$B\ga B_Q=B_0/\alpha^2=4.414\times 10^{13}$~G\@. 
This is valid for two reasons: 
(i) The free-electron energy in relativistic theory is 
\be
E=\left[c^2p_z^2+m_e^2c^4\left(1+2n_L {B\over B_Q}\right)\right]^{1/2}.
\label{eqrel}
\ee
For electrons in the ground Landau level ($n_L=0$), Eq.~(\ref{eqrel})
reduces to $E\simeq m_ec^2+p_z^2/(2m_e)$ for $p_zc\ll m_ec^2$; the
electron remains non-relativistic in the $z$ direction as long as the
electron energy is much less than $m_ec^2$;
(ii) it is well known \citep[e.g.,][]{SokTer} that
Eq.~(\ref{Wmeq}) describes the transverse motion of
an electron with $n_L=0$ at any field strength,
and thus Eq.~(\ref{Wmeq}) is valid in
the relativistic theory. Our calculations assume that the longitudinal
motion of the electron is non-relativistic. This is valid for helium at
all field strengths considered in this paper. Thus relativistic
corrections to our calculated electron wave functions, binding
energies, and transition cross sections are all small. Our
approximation is justified in part by \citet{chen92}, who find that
the relativistic corrections to the binding energy of the hydrogen
atom are of order $\Delta E/E \sim 10^{-5.5}-10^{-4.5}$ for the range of
field strengths we are considering in this work
($B=10^{12}-10^{14}$~G).

A bound state of the He atom, in which one electron occupies the
$(m_1\nu_1)$ orbital, and the other occupies the $(m_2\nu_2)$ orbital,
is denoted by
$\ket{m_1\nu_1,m_2\nu_2}=\ket{W_{m_1}f_{m_1\nu_1},W_{m_2}f_{m_2\nu_2}}$
(clearly, $\ket{m_1\nu_1,m_2\nu_2}=\ket{m_2\nu_2,m_1\nu_1}$).
The two-electron wave function is
\ba&&\hspace*{-2em}
\Psi_{m_1\nu_1,m_2\nu_2}(\bm{r}_1,\bm{r}_2)
 = \frac{1}{\sqrt{2}} \big[ W_{m_1}(\bm{r}_{1\perp})f_{m_1\nu_1}(z_1)
\nonumber\\&&\qquad\times
  W_{m_2}(\bm{r}_{2\perp})f_{m_2\nu_2}(z_2)
\nonumber\\&&
 - W_{m_2}(\bm{r}_{1\perp})f_{m_2\nu_2}(z_1)
  \, W_{m_1}(\bm{r}_{2\perp})f_{m_1\nu_1}(z_2) \big] \,.
\ea

The one-electron wave functions are found using Hartree--Fock theory, by
varying the total energy with respect to the wave functions. The total
energy is given by (see, e.g., \citealt{neuhauser87}):
\be
E = E_K + E_{eZ} + E_{\rm dir} + E_{\rm exc} \,,
\label{energyeq}
\ee
where
\ba&&\hspace*{-2em}
E_K = \frac{\hbar^2}{2m_e} \sum_{m\nu} \int \dd z \, |f'_{m\nu}(z)|^2 \,,
\\&&\hspace*{-2em}
E_{eZ} = -Ze^2 \sum_{m\nu} \int \dd z \, |f_{m\nu}(z)|^2 V_m(z) \,,
\\&&\hspace*{-2em}
E_{\rm dir} = \frac{e^2}{2} \sum_{m\nu,m'\nu'} \int \!\! \int \dd z \dd z' \,
 |f_{m\nu}(z)|^2 \, |f_{m'\nu'}(z')|^2
\nonumber\\&&\qquad\qquad\qquad\times
  D_{mm'}(z-z') \,,
\\&&\hspace*{-2em}
E_{\rm exc} = -\frac{e^2}{2} \sum_{m\nu,m'\nu'} \int \!\! \int \dd z \dd z' \,
 f^*_{m'\nu'}(z) f_{m\nu}(z)
\nonumber\\&&\qquad\qquad\times
   f^*_{m\nu}(z') f_{m'\nu'}(z') E_{mm'}(z-z') \,;
\ea
and
\be
V_m(z) = \int \dd\bm{r}_\perp \, \frac{|W_m(\bm{r}_\perp)|^2}{\bm{r}} \,,
\ee
\be
D_{mm'}(z-z') = \int \!\! \int \dd\bm{r}_\perp \dd\bm{r'}_\perp \,
 \frac{|W_m(\bm{r}_\perp)|^2 |W_{m'}(\bm{r'}_\perp)|^2}{|\bm{r'} - \bm{r}|} \,,
\ee
\ba&&\hspace*{-2em}
E_{mm'}(z-z') = \int \!\! \int \dd\bm{r}_\perp \dd\bm{r'}_\perp \,
 \frac{1}{|\bm{r'} - \bm{r}|}
\nonumber\\&&\qquad\qquad\times
    W^*_{m'}(\bm{r}_\perp) W_m(\bm{r}_\perp)
  W^*_m(\bm{r'}_\perp) W_{m'}(\bm{r'}_\perp) \,.
\ea
Variation of Eq.~(\ref{energyeq}) with respect to $f_{m\nu}(z)$ yields
\ba&&\hspace*{-2em}
\bigg[ -\frac{\hbar^2}{2m_e}\frac{\dd^2}{\dd z^2} - Ze^2 V_m(z)
\nonumber\\&&\hspace*{-1em}
     + e^2 \sum_{m'\nu'} \int \dd z' \, |f_{m'\nu'}(z')|^2 D_{mm'}(z-z')
      - \varepsilon_{m\nu} \bigg] f_{m\nu}(z)
\nonumber\\&&\hspace*{-1em}
    = e^2 \sum_{m'\nu'} \int \dd z' \,
     f^*_{m\nu}(z') f_{m'\nu'}(z') E_{mm'}(z-z') f_{m'\nu'}(z) \,.
\nonumber\\&&
    ~~
\label{HFeqs}
\ea
In these equations, asterisks denote complex conjugates, and
$f'_{m\nu}(z)\equiv \dd f_{m\nu}/\dd z$.  The wave functions
$f_{m\nu}(z)$ must satisfy appropriate boundary conditions, i.e.,
$f_{m\nu} \rightarrow 0$ as $z \rightarrow \pm\infty$, and must have
the required symmetry [$f_{m\nu}(z)=\pm f_{m\nu}(-z)$] and
the required number of nodes ($\nu$). The equations are solved
iteratively until self-consistency
is reached for each wave function $f_{m\nu}$ and energy
$\varepsilon_{m\nu}$. The total energy of the bound He state $\ket{m_1\nu_1,m_2\nu_2}$ can then be found, using either Eq.~(\ref{energyeq}) or
\be
E = \sum_{m\nu} \varepsilon_{m\nu} - E_{\rm dir} - E_{\rm exc} \,.
\label{energyeq2}
\ee

\subsection{Continuum states of the helium atom}
\label{sect:continuum}

The He state in which one electron occupies the bound $(m_3\nu_3)$
orbital, and other occupies the continuum state $(m_4k)$ is denoted by
$\ket{m_3\nu_3,m_4k}=\ket{W_{m_3}f_{m_3\nu_3},W_{m_4}f_{m_4k}}$. The
corresponding two-electron wave function is
\ba&&\hspace*{-2em}
\Psi_{m_3\nu_3,m_4k}(\bm{r}_1,\bm{r}_2) = \frac{1}{\sqrt{2}}
[ W_{m_3}(\bm{r}_{1\perp})f_{m_3\nu_3}(z_1)
\nonumber\\&&\qquad\times
W_{m_4}(\bm{r}_{2\perp})f_{m_4k}(z_2) \nonumber\\
 & & \qquad - W_{m_4}(\bm{r}_{1\perp})f_{m_4k}(z_1)
W_{m_3}(\bm{r}_{2\perp})f_{m_3\nu_3}(z_2) ] \,.
\ea
Here $f_{m_4k}(z)$ is the longitudinal wave function of the continuum
electron, and $k$ is the $z$-wavenumber of the electron at
$|z|\rightarrow\infty$ (far away from the He nucleus).

We can use Hartree--Fock theory to solve for the ionized He states as
we did for the bound He states. Since the continuum electron
wave function $f_{m_4k}(z)$ is non-localized in $z$, while the bound
electron wave function $f_{m_3\nu_3}(z)$ is localized around $z=0$, it
is a good approximation to neglect the continuum electron's influence
on the bound electron. We therefore solve for the bound electron
orbital using the equation
\be
\left[ -\frac{\hbar^2}{2m_e}\frac{d^2}{\dd z^2} - Ze^2 V_{m_3}(z)
\right] f_{m_3\nu_3}(z) = \varepsilon_{m_3\nu_3} f_{m_3\nu_3}(z)
\,.
\ee
The continuum electron, however, is influenced by the bound electron,
and its longitudinal wave function is determined from
\ba&&\hspace*{-2em}
\bigg[ -\frac{\hbar^2}{2m_e}\frac{\dd^2}{\dd z^2} - Ze^2 V_{m_4}(z)
\nonumber\\&&\hspace*{-1em}
     + e^2 \int \dd z' \, |f_{m_3\nu_3}(z')|^2 D_{m_3m_4}(z-z')
      - \varepsilon_f \bigg] f_{m_4k}(z)
\nonumber\\&&\hspace*{-1em}
    = e^2 \int \dd z' \,
     f^*_{m_4k}(z') f_{m_3\nu_3}(z') E_{m_3m_4}(z-z') f_{m_3\nu_4}(z) \,.
\nonumber\\&&
    ~~
\label{eq:cont}
\ea
where $\varepsilon_f=\varepsilon_{m_4k}=\hbar^2k^2/(2m_e)$.  Here, the
bound electron orbital $\ket{m_3\nu_3}$ satisfies the same boundary
conditions as discussed in Sect.\,\ref{sect:bound}. The shape of the
free electron wave function is determined by the energy of the
incoming photon and the direction the electron is emitted from the
ion. We will discuss this boundary condition in the next section.
The total energy of the ionized He state $\ket{m_3\nu_3,m_4k}$ is simply
\be
E = \varepsilon_{m_3\nu_3}+\varepsilon_f \,.
\label{energyeq3}
\ee
Note that the correction terms $E_{\rm dir}$ and $E_{\rm exc}$ that
appear in Eq.~(\ref{energyeq2}) do not also appear in
Eq.~(\ref{energyeq3}). The direct and exchange energies depend on the
local overlap of the electron wave functions, but the non-localized
nature of the free electron ensures that these terms are zero for the
continuum states.

\section{Radiative transitions}
\label{sec:transitions}

We will be considering transitions of helium atoms from two initial
states: the ground state, $\ket{00,10}$, and the first excited state,
$\ket{00,20}$.

In the approximation of an infinitely massive, pointlike nucleus,
the Hamiltonian of the He atom in electromagnetic field 
is \citep[see, e.g.,][]{landau77}
\be
   H = \!\! \sum_{j=1,2}  \! \frac{1}{2m_e} \left(\bm{p}_j +
   \frac{e}{c}\bm{A}_\mathrm{tot}(\bm{r}_j)\right)^2 
   \!\!   -  \!\! \sum_{j=1,2} \!\!  \frac{2 e^2}{r_j^2}
   + \frac{e^2}{|\bm{r}_1 - \bm{r}_2|} ,
\ee
where $\bm{p}_j=-\mathrm{i}\hbar\nabla_j$ is the canonical momentum
operator, acting on the $j$th electron, $\bm{r}_j$ is the $j$th
electron radius vector, measured from the nucleus, and
$\bm{A}_\mathrm{tot}(\bm{r})$ is the vector potential of the field. In
our case, $\bm{A}_\mathrm{tot}(\bm{r}) = \bm{A}_B(\bm{r}) +
\bm{A}_\mathrm{em}(\bm{r})$, where $\bm{A}_B(\bm{r})$ and
$\bm{A}_\mathrm{em}(\bm{r})$ are vector potentials of the stationary
magnetic field and electromagnetic wave, respectively. The
interaction operator is $H_\mathrm{int} = H - H_0$, where $H_0$ is
obtained from $H$ by setting $\bm{A}_\mathrm{em}(\bm{r})=0$.  The
unperturbed Hamiltonian $H_0$ is responsible for the stationary states
of He, discussed in Sect.\,\ref{sect:stat}. The vector potential and
the wave functions may be subject to gauge transformations; the wave
functions presented in Sect.\,\ref{sect:stat} correspond to the
cylindrical gauge $\bm{A}_B(\bm{r})=\frac12 \bm{B}\times\bm{r}$.
Neglecting non-linear (quadratic in $A_\mathrm{em}$) term, we have
\be
   H_\mathrm{int} \approx
   \frac{e}{2m_e c}  \sum_{j=1,2} [\bm{\pi}_j\cdot
    \bm{A}_\mathrm{em}(\bm{r}_j) +
   \bm{A}_\mathrm{em}(\bm{r}_j) \cdot\bm{\pi}_j],
\ee
where
\be
   \bm{\pi} = \bm{p} + \frac{e}{c}\bm{A}_B(\bm{r}).
\ee
is the non-perturbed kinetic momentum operator: $\bm{\pi} = m_e
\dot{\bm{r}} = m_e (\mathrm{i}/\hbar) [H_0\,\bm{r} - \bm{r}\,H_0]$.

For a monochromatic wave of the form
$\bm{A}_\mathrm{em}(\bm{r}) \propto
\bm{\epsilon}\,\mathrm{e}^{\mathrm{i}\bm{q}\cdot\bm{r}}$,
where $\bm{\epsilon}$ is the unit polarization vector,
applying the Fermi's Golden Rule
and assuming the transverse polarization
($\bm{\epsilon}\cdot\bm{q} = 0$), one obtains the
following general formula for the 
cross section of absorption of radiation
from a given initial state $\ket{a}$
\citep[see, e.g.,][]{armstrong72}:
\be
   \sigma(\omega,\bm{\epsilon}) = \sum_b \frac{4\pi^2}{\omega c}
   \left|\bm{\epsilon}\cdot\bra{b}
   \mathrm{e}^{\mathrm{i}\bm{q}\cdot\bm{r}} \bm{j} \ket{a}
   \right|^2\,\delta(\omega - \omega_{ba}),
\label{eq:sigma}
\ee
where $\ket{b}$ is the final state,
$\omega=qc$ is the photon frequency,
$\omega_{ba} = (E_b - E_a)/\hbar$,
and $\bm{j}$ is the electric current operator.
In our case,
$
   \bm{j} = (-e/m_e) (\bm{\pi}_1 + \bm{\pi}_2).
$

We shall calculate the cross sections 
in the dipole approximation -- i.e., drop
$\mathrm{e}^{\mathrm{i}\bm{q}\cdot\bm{r}}$ from \req{eq:sigma}.
This approximation is sufficiently accurate for calculation of
the total cross section as long as $\hbar\omega\ll m_e c^2$
(cf., e.g., \citealt{PP93,PP97} for the case of H atom).
In the dipole approximation, \req{eq:sigma}
can be written as
\be
      \sigma(\omega,\bm{\epsilon}) = \sum_b \frac{2\pi^2 e^2}{m_e
      c} f_{ba} \delta(\omega - \omega_{ba}),
\label{eq:sigma1}
\ee
where
\be
f_{ba}=
{2\over \hbar\omega_{ba} m_e}\left|\bra{b} \bm{\epsilon} \cdot 
\bm{\pi}\ket{a}\right|^2 
={2m_e\omega_{ba}\over\hbar}\left|\bra{b} \bm{\epsilon} \cdot 
\bm{r}\ket{a}\right|^2
\label{vel/length}
\ee
is the oscillator strength.
In the second equality we have passed from the `velocity
form' to the `length form' of the matrix
element
\citep[cf., e.g.,][]{Chandra45}. 
These representations are identical for the exact wave functions,
but it is not so for approximate ones.
In the adiabatic approximation, the length representation
[i.e., the right-hand side of \req{vel/length}] is preferable
\citep*[see][]{PP93,PPV}.

To evaluate the matrix element, we decompose the unit polarization vector
$\bm{\epsilon}$ into three cyclic components,  
\be
\bmath{\epsilon} = \epsilon_- \hat{\bm{e}}_+ 
+ \epsilon_+ \hat{\bm{e}}_- + \epsilon_0 \hat{\bm{e}}_0,
\ee
with $\hat{\bm{e}}_0=\hat{\bm{e}}_z$ 
along the external magnetic field direction (the z-axis), 
$\hat{\bm{e}}_\pm =(\hat{\bm{e}}_x \pm \mathrm{i}\hat{\bm{e}}_y)
/\sqrt{2}$, and $\epsilon_\alpha=\hat{\bm{e}}_\alpha\cdot
\bmath\epsilon$ (with $\alpha=\pm,0$). Then we can write
the cross section as the sum of three components,
\be
\sigma(\omega,\bmath\epsilon)=
\sigma_+(\omega)|{\epsilon}_+|^2 + \sigma_{-}(\omega)|{\epsilon}_-|^2 
+\sigma_0(\omega) |{\epsilon}_0|^2,
\label{3comp}
\ee
where $\sigma_\alpha$ has the same form as Eq.~(\ref{eq:sigma1}), with 
the corresponding oscillator strength given by
\be
f^{\alpha}_{ba}
={2m_e\omega_{ba}\rho_0^2\over\hbar}|M_{ba}|^2
=\frac{2\omega_{ba}}{\omc}|M_{ba}|^2,
\label{oscstreq}
\ee
with
\be
M_{ba}=\bra{b} \hat{\bm{e}}_\alpha^\ast\cdot\bar{\bm{r}}\ket{a},
\ee
where $\bar{\bm{r}}=\bm{r}/\rho_0$
and $\omc=eB/(m_e c)$ is the electron cyclotron frequency.

\subsection{Bound-bound transitions}
\label{sect:bb}

Consider the electronic transition 
\ba
&&\hspace*{-2em}
\ket{a}=\ket{m\nu,m_2\nu_2}=
\ket{W_mf_{m\nu},W_{m_2}f_{m_2\nu_2}}\nonumber\\
&&
\longrightarrow
\ket{b}=\ket{m'\nu',m_2\nu_2} 
=\ket{W_{m'}g_{m'\nu'},W_{m_2}g_{m_2\nu_2}}.
\label{bbtranseq}
\ea
The selection rules for allowed transitions
and the related matrix elements are
\ba
\label{seleq1}
&&\hspace*{-2em}
\sigma_0:\quad \Delta m=0,~\Delta\nu=\mbox{odd},
\nonumber\\&&
\qquad\quad M_{ba}=
\langle g_{m\nu'}|{\bar z}|f_{m\nu}\rangle
\langle g_{m_2\nu_2}|f_{m_2\nu_2}\rangle,\\
&&\hspace*{-2em}
\label{seleq2}
\sigma_+:\quad \Delta m=1,~\Delta\nu=\mbox{even},
\nonumber\\&&
\qquad\quad M_{ba}=
\sqrt{m+1}\,\langle g_{m'\nu'}|f_{m\nu}\rangle
\langle g_{m_2\nu_2}|f_{m_2\nu_2}\rangle,\\
&&\hspace*{-2em}
\sigma_-:\quad \Delta m=-1,~\Delta\nu=\mbox{even},
\nonumber\\&&
\qquad\quad
 M_{ba}=
\sqrt{m}\,\langle g_{m'\nu'}|f_{m\nu}\rangle
\langle g_{m_2\nu_2}|f_{m_2\nu_2}\rangle,
\label{seleq3}
\ea
where $\Delta m=m'-m,~\Delta\nu=\nu'-\nu$.
The oscillator strengths for bound-bound transitions from
the states $\ket{00,10}$ and $\ket{00,20}$ 
are given in Table~\ref{osctable}.

The selection rules (\ref{seleq1})\,--\,(\ref{seleq3})
are exact in the dipole approximation.
The selection rules in $m$ follow from the conservation
of the $z$-projection of total (for the photon and two electrons)
angular momentum. Technically, in the adiabatic
approximation, they follow from the properties
of the Landau functions \citep[e.g.,][]{PP93}.
The selection rules in $\nu$ follow from the fact that the
functions $g_{m'\nu'}$ and $f_{m\nu}$ 
have the same parity for even $\nu'-\nu$
and opposite parity for odd $\nu'-\nu$.

In addition to these selection rules, there are
approximate selection rules which rely on
the approximate orthogonality of functions
$g_{m'\nu'}$ and $f_{m\nu}$ (for general $\nu\ne\nu'$).
Because of this approximate orthogonality,
which holds better the larger $B$ is,
we have 
\be
   \langle g_{m'\nu'} | f_{m\nu} \rangle \langle g_{m_2\nu_2} |
   f_{m_2\nu_2} \rangle = \delta_{\nu,\nu'} + \varepsilon,
\label{ortho}
\ee
where $|\varepsilon| \ll 1$ and $\varepsilon \rightarrow 0$ as $\Delta\nu \rightarrow \pm\infty$.
Therefore,
the oscillator strengths for transitions
with $\alpha=\pm$ and $\Delta\nu=2,4,\ldots$
are small compared to those
with $\Delta\nu=0$.
 The latter oscillator strengths can be approximated, according to
 Eqs.~(\ref{oscstreq}), (\ref{seleq2}), (\ref{seleq3}) and
 (\ref{ortho}), by
\be
   f^+_{ba} \approx 2(m+1)\, \omega_{ba}/\omc ,
   \qquad
   f^-_{ba} \approx 2m\, \omega_{ba}/\omc
\label{f_approx}
\ee
($\alpha=\Delta m=\pm1$, $\nu'=\nu$).

The same approximate orthogonality leads to the smallness
of matrix elements for transitions
of the type
$\ket{m\nu,m_2\nu_2} \longrightarrow
   \ket{m'\nu',m_2\nu_2'}$
with $\nu_2'\neq\nu_2$ for $\alpha=\pm$
and the smallness of cross terms in the matrix elements of the form
$\langle g_{m_2\nu_2} | f_{m\nu} \rangle \langle g_{m'\nu'} |
   f_{m_2\nu_2} \rangle$ when $m' = m_2$
(i.e., the so-called ``one-electron jump rule'');
we have therefore excluded such terms from the selection rule equations
above [Eqs.~(\ref{seleq1})\,--\,(\ref{seleq3})].

\subsection{Photoionization}
\label{sect:photo}

The bound-free absorption cross section for the transition from the bound state
$\ket{b}$ to the continuum state $\ket{f}$ is
given by \req{eq:sigma} with obvious substitutions
$\ket{a}\to\ket{b}$, $\ket{b}\to\ket{f}$,
and 
\be
\sum_f \rightarrow (L_z/2\pi)\int_{-\infty}^\infty \dd k,
\ee
where $L_z$ is the normalization length of the continuum electron
[$\int_{-L_z/2}^{L_z/2} \dd z \, |g_{mk}(z)|^2 = 1$]
and $k$ is the wave number of the
outgoing electron (Sect.\,\ref{sect:continuum}). 
Therefore we have
\ba&&\hspace*{-2em}
\sigma_\mathrm{bf}(\omega,\bm{\epsilon})
 = \frac{2\pi e^2 L_z}{m_e c\hbar^2 \omega_{fb} k} 
\Big\{
\left|\bra{f_{k}} \mathrm{e}^{\mathrm{i}\bm{q} \cdot \bm{r}}
\bm{\epsilon} \cdot \bm{\pi}\ket{b}\right|^2
\nonumber\\&&\qquad\qquad
 +
\left|\bra{f_{-k}} \mathrm{e}^{\mathrm{i}\bm{q} \cdot \bm{r}}
\bm{\epsilon} \cdot \bm{\pi}\ket{b}\right|^2
\Big\} \,,
\label{bfxseceq}
\ea
where $k = \sqrt{2 m_e \varepsilon_f}/\hbar$ and
$\ket{f_{\pm k}}$ 
represents the final state where the free electron 
has wave number $\pm k$ (here and hereafter we assume $k > 0$). 
The asymptotic conditions for these outgoing 
free electrons are 
\citep[cf., e.g.,][]{PPV}
$
  g_{mk}(z) \sim \exp[\mathrm{i} \varphi_k(z)]
$
at $z\to\pm\infty$,
where
$
   \varphi_k(z) = |kz| + (k a_0)^{-1} \ln|kz|
$
and $a_0=\hbar^2/m_e e^2$ is the Bohr radius.
Since we do not care about direction of the outgoing electron,
we can use for calculations a basis of symmetric
and antisymmetric wave functions 
of the continuum -- that is, in \req{bfxseceq} we can replace
$\bra{f_{k}}$ and $\bra{f_{-k}}$ by 
$\bra{f_\mathrm{even}}$ and $\bra{f_\mathrm{odd}}$.
The symmetric state $\ket{f_{\rm even}}$ is determined by 
the free electron boundary condition $g'_{mk,\mathrm{even}}(0)=0$
and
the antisymmetric state $\ket{f_{\rm odd}}$ is determined 
by $g_{mk,\mathrm{odd}}(0)=0$.
Since the coefficients in \req{eq:cont} are real, 
 $g_{mk,\mathrm{even}}(z)$ and  $g_{mk,\mathrm{odd}}(z)$
can be chosen real.
At $z\to\pm\infty$, they behave as
$g_{mk,\mathrm{(even,odd)}}(z)\sim\sin[\varphi(z)+\mathrm{constant}]$
(where the value of constant depends on all quantum numbers,
including $k$).
We still have the normalization 
$\int_{-L_z/2}^{L_z/2} \dd z \, |g_{mk,\mathrm{(even,odd)}}(z)|^2 = 1$. 

Similar to bound-bound transitions, we can decompose the bound-free
cross section into three components,
\req{3comp}.
Thus, using the dipole approximation and the length form of the
matrix elements, as discussed above, we have
for ($\alpha=\pm,0$)-components of the bound-free cross section
\ba&&\hspace*{-2em}
\sigma_{\mathrm{bf},\alpha}(\omega) 
 = \frac{3}{4}\, \sigma_{\mathrm{Th}}
 \left( \frac{m_e c^2}{\hbar \omega} \right)^3
 \sqrt{\frac{m_e c^2}{2\varepsilon_f}} \left( \frac{L_z a_0}{\rho_0^2}
 \right) 
 \left(\frac{\omega \rho_0}{c}\right)^4
\nonumber\\&&\qquad\qquad\qquad\times
\left| \bra{f} 
 \hat{\bm{e}}_\alpha^\ast\cdot\bar{\bm{r}}\ket{b} \right|^2,
\label{xsecteq}
\ea
where $\ket{f}=\ket{f_\mathrm{even}}$ or $\ket{f}=\ket{f_\mathrm{odd}}$
depending on the parity of the initial state and
according to the selection rules,
and
$
\sigma_{\mathrm{Th}}=(8\pi/3)\,({e^2}/{m_e c^2})^2
$
is the Thomson cross section.
The selection rules and related matrix elements 
for the bound-free transitions 
\ba
&&\ket{b}=\ket{m\nu,m_2\nu_2}=
\ket{W_mf_{m\nu},W_{m_2}f_{m_2\nu_2}}\nonumber\\
&&\quad\longrightarrow
\ket{f}=\ket{m'k,m_2\nu_2} 
=\ket{W_{m'}g_{m'k},W_{m_2}g_{m_2\nu_2}}
\ea
are similar to those for the bound-bound transitions
[see Eqs.~(\ref{seleq1})\,--\,(\ref{seleq3})]:
\ba
\label{Fseleq1}
&&\hspace*{-2em}
\sigma_0:\quad \Delta m=0,~\Delta\nu=\mbox{odd},
\nonumber\\&&
\qquad\quad M_{fb}=
\langle g_{mk}|{\bar z}|f_{m\nu}\rangle
\langle g_{m_2\nu_2}|f_{m_2\nu_2}\rangle,\\
&&\hspace*{-2em}
\label{Fseleq2}
\sigma_+:\quad \Delta m=1,~\Delta\nu=\mbox{even},
\nonumber\\&&
\qquad\quad M_{fb}=
\sqrt{m+1}\,\left(\langle g_{m'k}|f_{m\nu}\rangle
\langle g_{m_2\nu_2}|f_{m_2\nu_2}\rangle\right.
\nonumber\\&&
\qquad\qquad \left. -\delta_{m'\nu,m_2\nu_2}
\langle g_{m_2\nu_2}|f_{m\nu}\rangle
\langle g_{m'k}|f_{m_2\nu_2}\rangle\right),\\
&&\hspace*{-2em}
\sigma_-:\quad \Delta m=-1,~\Delta\nu=\mbox{even},
\nonumber\\&&
\qquad\quad
 M_{fb}=
\sqrt{m}\,\left(\langle g_{m'k}|f_{m\nu}\rangle
\langle g_{m_2\nu_2}|f_{m_2\nu_2}\rangle\right.
\nonumber\\&&
\qquad\qquad \left. -\delta_{m'\nu,m_2\nu_2}
\langle g_{m_2\nu_2}|f_{m\nu}\rangle
\langle g_{m'k}|f_{m_2\nu_2}\rangle\right),
\label{Fseleq3}
\ea
In this case, the condition $\Delta\nu=\mbox{odd}$ means that 
$g_{m'k}$ and $f_{m\nu}$ must have opposite parity, 
and the condition $\Delta\nu=\mbox{even}$ means that 
$g_{m'k}$ and $f_{m\nu}$ must have the same parity.
The oscillator strengths for bound-free transitions from
the states $\ket{00,10}$ and $\ket{00,20}$ 
are given in Table~\ref{fittable}.

\begin{figure*}
\begin{center}
\begin{tabular}{cc}
\resizebox{3in}{!}{\includegraphics{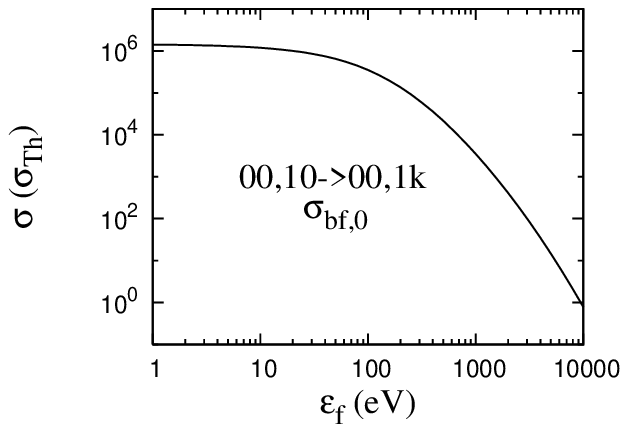}} &
\resizebox{3in}{!}{\includegraphics{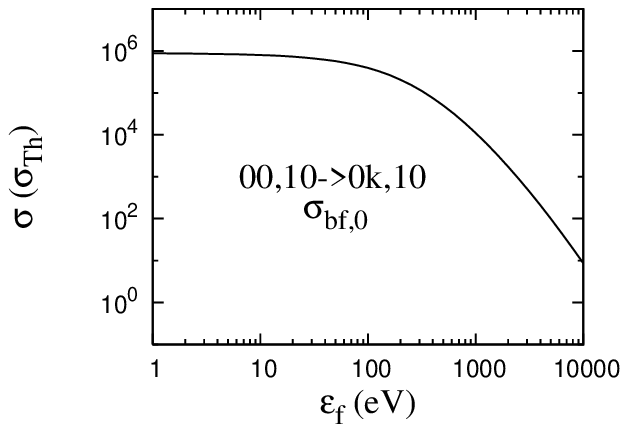}} \\
\resizebox{3in}{!}{\includegraphics{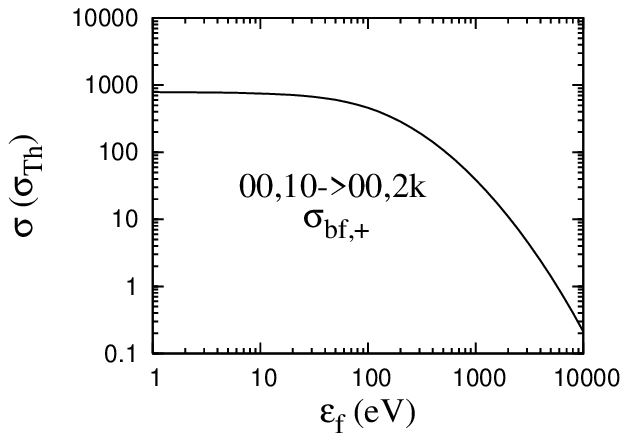}} &
\resizebox{3in}{!}{\includegraphics{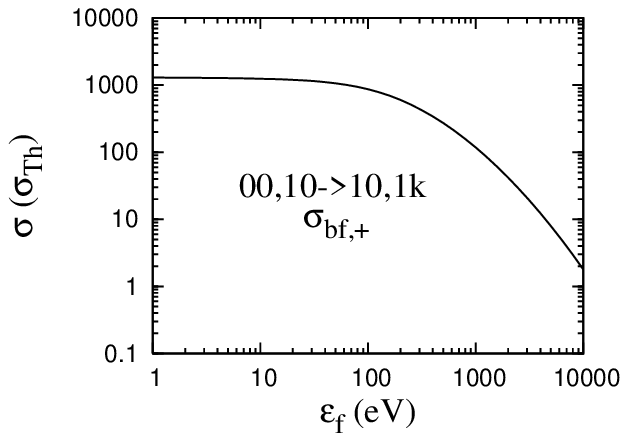}} \\
\resizebox{3in}{!}{\includegraphics{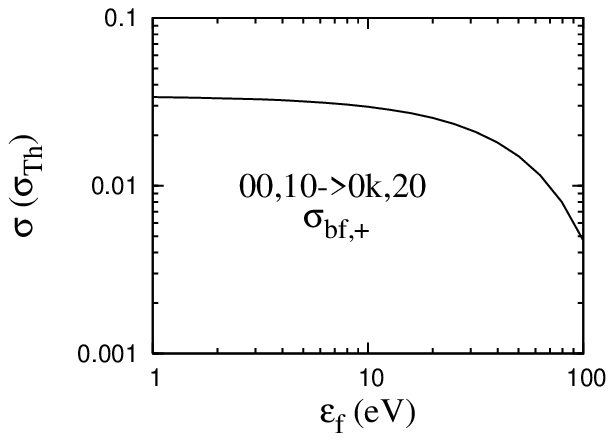}} &
\resizebox{3in}{!}{\includegraphics{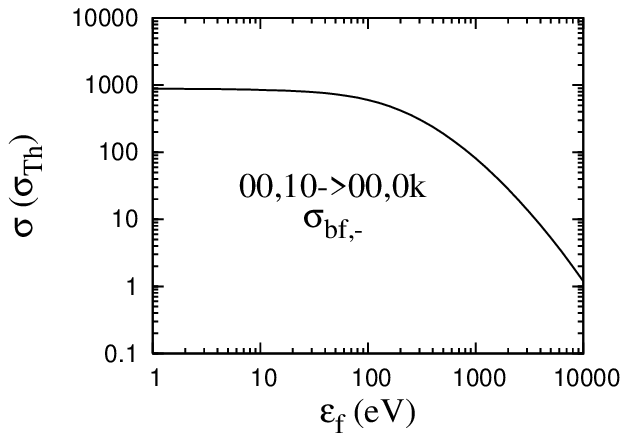}} \\
\end{tabular}
\caption{Partial cross sections $\sigma_{(0,+,-)}$ versus final 
ionized electron energy for photoionization of the ground state 
helium atom ($(m_1,m_2)=(1,0)$). The field strength is $10^{12}$ G. 
The transition $\ket{00,10}\rightarrow\ket{0k,20}$ in the bottom 
left panel is an example of a `weak' transition. 
We have ignored these transitions in our calculations
of the total cross sections.}
\label{partialfig}
\end{center}
\end{figure*}

Note that in Eqs.~(\ref{Fseleq2}) and (\ref{Fseleq3}), the second
term in the matrix element (of the form $\langle
g_{m_2\nu_2}|f_{m\nu}\rangle \langle g_{m'k}|f_{m_2\nu_2} \rangle$)
corresponds to transitions of both electrons. This appears to violate
the ``one-electron jump rule'' and other approximate selection rules
discussed in Sect.\,\ref{sect:bb} [see Eq.~(\ref{ortho})]. In fact,
these approximate rules are not directly relevant for bound-free
transitions, since the matrix elements involving a continuuum state
are always small: $\langle g_{m'k}|f_{m\nu}\rangle \rightarrow 0$ as
the normalization length $L_z \rightarrow \infty$.  Rather, we use a
different set of selection rules to determine which of these `small'
matrix elements are smaller than the rest.  The first is that
\ba&&\hspace*{-2em}
   \langle g_{m'k} | f_{m\nu} \rangle
   \langle g_{m_2\nu_2} | f_{m_2\nu_2} \rangle \gg
   \langle g_{m'k} | f_{m\nu} \rangle
   \langle g_{m_2\nu_2'} | f_{m_2\nu_2} \rangle,
   \nonumber\\&&
\ea
when $\nu_2' \ne \nu_2$.
This selection rule is similar to the bound-bound transition case
as $\langle g_{m_2\nu_2'} | f_{m_2\nu_2} \rangle$ involves a
bound electron transition, not a free electron
transition. The second approximate selection rule that applies here
is more complicated: terms of the form
$\langle g_{m'\nu} | f_{m\nu} \rangle
\langle g_{m_2k} | f_{m_2\nu_2} \rangle$ are small, unless $m' = m_2$
and $\nu_2 = \nu$. This exception for $m' = m_2$ and $\nu_2 = \nu$
is due to the exchange term in the differential equation for the
free electron wave function [Eq.~(\ref{eq:cont})], which strongly
(anti)correlates the two final wave functions $\ket{g_{m'\nu}}$
and $\ket{g_{m_2k}}$. If $m' = m_2$ and $\nu = \nu_2$, then since
$\langle g_{m'\nu} | f_{m_2\nu_2} \rangle$ is not small (in fact, it
is of order 1), $\langle g_{m_2k} | f_{m_2\nu_2} \rangle$ will not
be small but will be of the same order as other terms involving
the free electron wave function. In particular,
the second selection rule means, e.g., that the matrix element for
the transition from $\ket{00,10}$ to $\ket{00,0k}$ is
\be
   M_{00,10\rightarrow00,0k} =
   \langle g_{0k} | f_{10} \rangle \langle g_{00} | f_{00} \rangle
   - \langle g_{00} | f_{10} \rangle \langle g_{0k} | f_{00} \rangle,
\label{0k010transeq}
\ee
where the second term is non-negligible, but that the matrix element
for the transition from $\ket{00,10}$ to $\ket{0k,20}$, which is
\be
   M_{00,10\rightarrow0k,20} =
   \langle g_{20} | f_{10} \rangle \langle g_{0k} | f_{00} \rangle,
\ee
is small compared to the other matrix elements and can be ignored
(see Fig.~\ref{partialfig}).

We make one final comment here about the effect of exchange
interaction on the free electron state. If the exchange term [the
right-hand side of Eq.~(\ref{eq:cont})] is neglected in the
calculation of the free electron wave function, then the cross terms
(i.e., those involving two-electron transitions) in the matrix
elements of Eqs.~(\ref{Fseleq2}) and (\ref{Fseleq3}) are small and can
be neglected. One then obtains approximate photoionization cross
sections which are within a factor of two of the true values in most
cases and much better for $\sigma_0$ transitions. If the exchange term
is included in Eq.~(\ref{eq:cont}) but the cross terms in the matrix
elements are ignored, significant errors in the $\sigma_\pm$
photoionization cross sections will result. To obtain reliable cross
sections for all cases, both the exchange effect on the free electron
and the contribution of two-electron transitions must be included.

\section{Results}
\label{sect:res}

\begin{table*}
\begin{minipage}{126mm}
\caption{Bound-bound transitions $\ket{a} \rightarrow \ket{b}$:
The photon energy $\hbar\omega_{ba}=E_b-E_a$ (in eV) and the
oscillator strength $f_{ba}^\alpha$ for different polarization
components $\alpha$ [see Eq.~(\ref{oscstreq})]. All transitions
$\Delta\nu \le 1$ from the initial states $\ket{00,10}$ and
$\ket{00,20}$ are listed, for several magnetic field strengths
$B_{12}=B/(10^{12}\mbox{ G})$.
The last two columns list the transition energies
$\hbar\omega_{ba}^*$ and oscillator strengths $f_{ba}^*$,
corrected for the finite mass of the nucleus, 
according to Sect.\,\ref{sect:correction0}.
}
\centering
\begin{tabular}{c @{\quad} c @{\quad} r l @{\quad} c @{\quad} c @{\quad} c @{\quad} c}
\hline\hline
$B_{12}$ & $\sigma$ & $\ket{a}$ & $\rightarrow \ket{b}$
 & $\hbar\omega_{ba}$ & $f_{ba}$ & $\hbar\omega_{ba}^*$ & $f_{ba}^*$ \\
\hline
1 & 0 & $\ket{00,10}$ & $\rightarrow \ket{00,11}$ & 147.5 & 0.234 & -- & -- \\
 & & & $\rightarrow \ket{10,01}$ & 271.8 & 0.124 & -- & -- \\
 & + & & $\rightarrow \ket{00,20}$ & 43.11 & 0.0147 & 44.70 & 0.0153 \\
\hline
 & 0 & $\ket{00,20}$ & $\rightarrow \ket{00,21}$ & 104.4 & 0.312 & -- & -- \\
 & & & $\rightarrow \ket{20,01}$ & 277.7 & 0.115 & -- & -- \\
 & + & & $\rightarrow \ket{00,30}$ & 18.01 & 0.00930 & 19.60 & 0.0101 \\
 & & & $\rightarrow \ket{20,10}$ & 100.7 & 0.0170 & 102.3 & 0.0172 \\
\hline\hline
5 & 0 & $\ket{00,10}$ & $\rightarrow \ket{00,11}$ & 256.2 & 0.127 & -- & -- \\
 & & & $\rightarrow \ket{10,01}$ & 444.8 & 0.0603 & -- & -- \\
 & + & & $\rightarrow \ket{00,20}$ & 66.95 & 0.00459 & 74.89 & 0.00512 \\
\hline
 & 0 & $\ket{00,20}$ & $\rightarrow \ket{00,21}$ & 189.2 & 0.176 & -- & -- \\
 & & & $\rightarrow \ket{20,01}$ & 455.0 & 0.0537 & -- & -- \\
 & + & & $\rightarrow \ket{00,30}$ & 28.94 & 0.00299 & 36.88 & 0.00381 \\
 & & & $\rightarrow \ket{20,10}$ & 151.1 & 0.00512 & 159.0 & 0.00539 \\
\hline\hline
10 & 0 & $\ket{00,10}$ & $\rightarrow \ket{00,11}$ & 318.9 & 0.0974 & -- & -- \\
 & & & $\rightarrow \ket{10,01}$ & 540.8 & 0.0457 & -- & -- \\
 & + & & $\rightarrow \ket{00,20}$ & 79.54 & 0.00273 & 95.42 & 0.00327 \\
\hline
 & 0 & $\ket{00,20}$ & $\rightarrow \ket{00,21}$ & 239.4 & 0.136 & -- & -- \\
 & & & $\rightarrow \ket{20,01}$ & 553.3 & 0.0405 & -- & -- \\
 & + & & $\rightarrow \ket{00,30}$ & 34.84 & 0.00179 & 50.72 & 0.00261 \\
 & & & $\rightarrow \ket{20,10}$ & 177.0 & 0.00301 & 192.9 & 0.00328 \\
\hline\hline
50 & 0 &  $\ket{00,10}$ & $\rightarrow \ket{00,11}$ & 510.9 & 0.0557 & -- & -- \\
 & & & $\rightarrow \ket{10,01}$ & 822.2 & 0.0266 & -- & -- \\
 & + & & $\rightarrow \ket{00,20}$ & 114.2 & 7.85e$-$4 & 193.6 & 0.00133 \\
\hline
 & 0 & $\ket{00,20}$ & $\rightarrow \ket{00,21}$ & 396.7 & 0.0776 & -- & -- \\
 & & & $\rightarrow \ket{20,01}$ & 841.1 & 0.0235 & -- & -- \\
 & + & & $\rightarrow \ket{00,30}$ & 51.92 & 5.37e$-$4 & 131.3 & 0.00136 \\
 & & & $\rightarrow \ket{20,10}$ & 246.5 & 8.41e$-$4 & 325.9 & 0.00111 \\
\hline\hline
100 & 0 & $\ket{00,10}$ & $\rightarrow \ket{00,11}$ & 616.4 & 0.0452 & -- & -- \\
 & & & $\rightarrow \ket{10,01}$ & 971.4 & 0.0221 & -- & -- \\
 & + & & $\rightarrow \ket{00,20}$ & 131.4 & 4.52e$-$4 & 290.2 & 9.98e$-$4 \\
\hline
 & 0 & $\ket{00,20}$ & $\rightarrow \ket{00,21}$ & 485.0 & 0.0626 & -- & -- \\
 & & & $\rightarrow \ket{20,01}$ & 993.4 & 0.0195 & -- & -- \\
 & + & & $\rightarrow \ket{00,30}$ & 60.57 & 3.13e$-$4 & 219.4 & 0.00114 \\
 & & & $\rightarrow \ket{20,10}$ & 280.7 & 4.80e$-$4 & 439.5 & 7.51e$-$4 \\
\hline\hline
\end{tabular}
\label{osctable}
\end{minipage}
\end{table*}

\begin{table*}
\begin{minipage}{126mm}
\caption{Bound-free transitions $\ket{b} \rightarrow \ket{f}$: The
threshold photon energy $\hbar\omega_{\rm thr}$ (in eV) and the
fitting parameters ${\cal A}$, ${\cal B}$, and ${\cal C}$ used in the
cross section fitting formulas [Eq.~(\ref{fitforms})]. All transitions
from the initial states $\ket{00,10}$ and $\ket{00,20}$ are listed,
for several magnetic field strengths $B_{12}=B/(10^{12}\mbox{ G})$.}
\centering
\begin{tabular}{c @{\quad} c @{\quad} r l @{\quad} c @{\quad} l @{\quad} r @{\qquad} l l}
\hline\hline
$B_{12}$ & $\sigma$ & $\ket{b}$ & $\rightarrow \ket{f}$ & $m_i$ & $\hbar\omega_{\rm thr}$ & ${\cal A}$~~ & ~~~${\cal B}$ & 
~~~${\cal C}$ \\
\hline
1 & 0 & $\ket{00,10}$ & $\rightarrow \ket{00,1k}$ & 1 & 159.2 & 0.96 & 0.093 & 1.43e6 \\
 & & & $\rightarrow \ket{10,0k}$ & 0 & 283.2 & 0.89 & 0.20 & 8.83e5 \\
 & + & & $\rightarrow \ket{00,2k}$ & 1 & 159.2 & 0.70 & 0.061 & 7.95e2\\
 & & & $\rightarrow \ket{10,1k}$ & 0 & 283.2 & 0.86 & 0.094 & 1.30e3 \\
 & -- & & $\rightarrow \ket{00,0k}$ & 1 & 159.2 & 0.62 & 0.030 & 8.89e2 \\
\hline
 & 0 & $\ket{00,20}$ & $\rightarrow \ket{00,2k}$ & 2 & 116.0 & 1.00 & 0.062 & 1.78e6 \\
 & & & $\rightarrow \ket{20,0k}$ & 0 & 289.2 & 0.88 & 0.22 & 8.71e5 \\
 & + & & $\rightarrow \ket{00,3k}$ & 2 & 116.0 & 0.66 & 0.038 & 3.94e2 \\
 & & & $\rightarrow \ket{20,1k}$ & 0 & 289.2 & 0.54 & 0.14 & 6.48e2 \\
 & -- & & $\rightarrow \ket{00,1k}$ & 2 & 116.0 & 0.62 & 0.029 & 5.82e2 \\
\hline\hline
5 & 0 & $\ket{00,10}$ & $\rightarrow \ket{00,1k}$ & 1 & 268.2 & 0.86 & 0.061 & 8.39e5 \\
 & & & $\rightarrow \ket{10,0k}$ & 0 & 456.4 & 0.69 & 0.16 & 4.60e5 \\
 & + & & $\rightarrow \ket{00,2k}$ & 1 & 268.2 & 0.68 & 0.036 & 1.14e2 \\
 & & & $\rightarrow \ket{10,1k}$ & 0 & 456.4 & 0.83 & 0.057 & 1.93e2 \\
 & -- & & $\rightarrow \ket{00,0k}$ & 1 & 268.2 & 0.60 & 0.020 & 1.36e2 \\
\hline
 & 0 & $\ket{00,20}$ & $\rightarrow \ket{00,2k}$ & 2 & 201.2 & 0.92 & 0.039 & 1.11e6 \\
 & & & $\rightarrow \ket{20,0k}$ & 0 & 466.5 & 0.65 & 0.18 & 4.39e5 \\
 & + & & $\rightarrow \ket{00,3k}$ & 2 & 201.2 & 0.65 & 0.021 & 5.95e1 \\
 & & & $\rightarrow \ket{20,1k}$ & 0 & 466.5 & 0.54 & 0.084 & 9.13e1 \\
 & -- & & $\rightarrow \ket{00,1k}$ & 2 & 201.2 & 0.61 & 0.015 & 7.82e1 \\
\hline\hline
10 & 0 & $\ket{00,10}$ & $\rightarrow \ket{00,1k}$ & 1 & 331.1 & 0.82 & 0.051 & 6.58e5 \\
 & & & $\rightarrow \ket{10,0k}$ & 0 & 552.5 & 0.63 & 0.15 & 3.51e5 \\
 & + & & $\rightarrow \ket{00,2k}$ & 1 & 331.1 & 0.67 & 0.029 & 4.94e1 \\
 & & & $\rightarrow \ket{10,1k}$ & 0 & 552.5 & 0.81 & 0.046 & 8.43e1 \\
 & -- & & $\rightarrow \ket{00,0k}$ & 1 & 331.1 & 0.59 & 0.016 & 6.00e1 \\
\hline
 & 0 & $\ket{00,20}$ & $\rightarrow \ket{00,2k}$ & 2 & 251.6 & 0.88 & 0.033 & 8.77e5 \\
 & & & $\rightarrow \ket{20,0k}$ & 0 & 564.9 & 0.59 & 0.16 & 3.31e5 \\
 & + & & $\rightarrow \ket{00,3k}$ & 2 & 251.6 & 0.64 & 0.017 & 2.64e1 \\
 & & & $\rightarrow \ket{20,1k}$ & 0 & 564.9 & 0.53 & 0.069 & 3.97e1 \\
 & -- & & $\rightarrow \ket{00,1k}$ & 2 & 251.6 & 0.61 & 0.012 & 3.25e1 \\
\hline\hline
50 & 0 & $\ket{00,10}$ & $\rightarrow \ket{00,1k}$ & 1 & 523.3 & 0.73 & 0.034 & 3.74e5 \\
 & & & $\rightarrow \ket{10,0k}$ & 0 & 834.2 & 0.54 & 0.11 & 1.96e5 \\
 & + & & $\rightarrow \ket{00,2k}$ & 1 & 523.3 & 0.63 & 0.020 & 7.15e0 \\
 & & & $\rightarrow \ket{10,1k}$ & 0 & 834.2 & 0.77 & 0.033 & 1.22e1 \\
 & -- & & $\rightarrow \ket{00,0k}$ & 1 & 523.3 & 0.57 & 0.012 & 8.94e0 \\
\hline
 & 0 & $\ket{00,20}$ & $\rightarrow \ket{00,2k}$ & 2 & 409.1 & 0.79 & 0.021 & 5.02e5 \\
 & & & $\rightarrow \ket{20,0k}$ & 0 & 853.0 & 0.50 & 0.13 & 1.83e5 \\
 & + & & $\rightarrow \ket{00,3k}$ & 2 & 409.1 & 0.62 & 0.0104 & 4.04e0 \\
 & & & $\rightarrow \ket{20,1k}$ & 0 & 853.0 & *0.52 & 0.052 & 5.88e0 \\
 & -- & & $\rightarrow \ket{00,1k}$ & 2 & 409.1 & 0.59 & 0.0058 & 4.13e0 \\
\hline\hline
100 & 0 & $\ket{00,10}$ & $\rightarrow \ket{00,1k}$ & 1 & 628.8 & 0.69 & 0.029 & 2.96e5 \\
 & & & $\rightarrow \ket{10,0k}$ & 0 & 983.4 & 0.51 & 0.101 & 1.56e5 \\
 & + & & $\rightarrow \ket{00,2k}$ & 1 & 628.8 & 0.62 & 0.019 & 3.12e0 \\
 & & & $\rightarrow \ket{10,1k}$ & 0 & 983.4 & 0.75 & 0.031 & 5.33e0 \\
 & -- & & $\rightarrow \ket{00,0k}$ & 1 & 628.8 & 0.56 & 0.012 & 3.94e0 \\
\hline
 & 0 & $\ket{00,20}$ & $\rightarrow \ket{00,2k}$ & 2 & 498.0 & 0.75 & 0.018 & 3.96e5 \\
 & & & $\rightarrow \ket{20,0k}$ & 0 & 1008 & 0.47 & 0.12 & 1.45e5 \\
 & + & & $\rightarrow \ket{00,3k}$ & 2 & 498.0 & 0.60 & 0.0092 & 1.81e0 \\
 & & & $\rightarrow \ket{20,1k}$ & 0 & 1008 & *0.50 & 0.050 & 2.60e0 \\
 & -- & & $\rightarrow \ket{00,1k}$ & 2 & 498.0 & 0.58 & 0.0042 & 1.69e0 \\
\hline\hline
\end{tabular}
\label{fittable}
\end{minipage}
\end{table*}

Tables~\ref{osctable} and \ref{fittable} give results for transitions 
of helium atoms from the ground state ($\ket{00,10}$) and the first 
excited state ($\ket{00,20}$). 
Table~\ref{osctable} gives results 
(photon energies and oscillator strengths) for all possible 
bound-bound transitions with $\Delta\nu \le 1$, for the field strengths 
$B_{12}=1,5,10,50,100$, 
where $B_{12}=B/(10^{12}\mbox{ G})$. 
Transitions $\ket{a}\to\ket{b}$ for $\alpha=-$
are not listed separately, being
equivalent to transitions $\ket{b}\to\ket{a}$ for $\alpha=+$.
One can check that the oscillator strengths $f_{ba}$ presented in 
Table~\ref{osctable} for $\alpha=+$ are well described
by the approximation (\ref{f_approx}).

\begin{figure}
\includegraphics[width=\columnwidth]{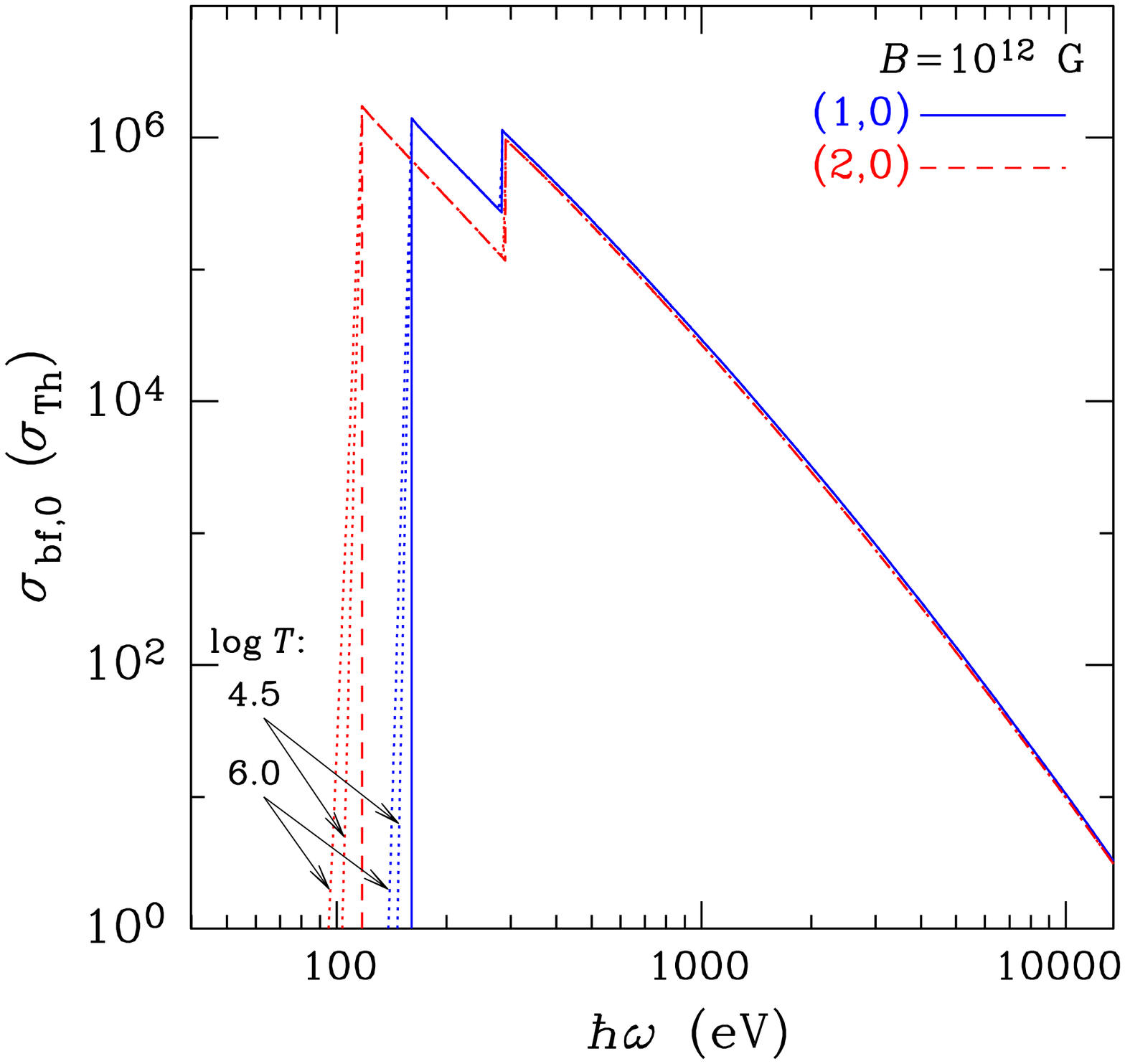}
\caption{Total cross section $\sigma_0$ versus photon energy for helium
photoionization, from initial states $(m_1,m_2)=(1,0)$ (solid lines)
and $(2,0)$ (dashed lines). The
field strength is $10^{12}$ G.
The dotted
lines extending from each cross section curve represent the
effect of magnetic broadening on these cross sections, as approximated
in Eq.~(\ref{magbroad}),
for $T=10^{4.5}$~K (steeper lines) and $10^6$~K (flatter lines).}
\label{totalfig}
\end{figure}

\begin{figure}
\includegraphics[width=\columnwidth]{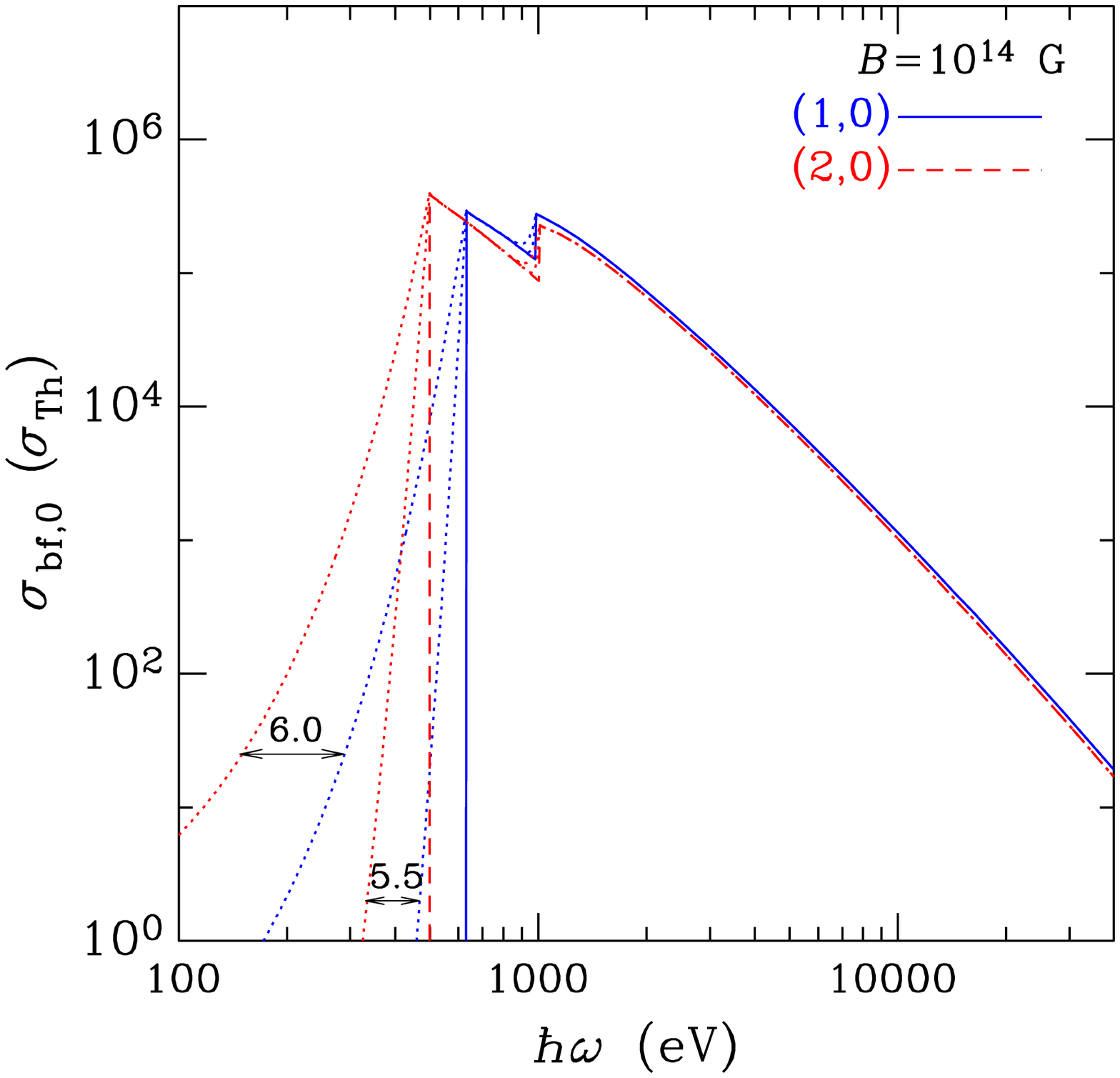}
\caption{Total cross section $\sigma_0$ versus photon energy for helium
photoionization, from initial states $(m_1,m_2)=(1,0)$ (solid lines)
 and $(2,0)$ (dashed lines). The
field strength is $10^{14}$ G.
The dotted
lines extending from each cross section curve represent the
effect of magnetic broadening on these cross sections, as approximated
in Eq.~(\ref{magbroad}),
for $T=10^{5.5}$~K (steeper lines) and $10^6$~K (flatter lines).}
\label{totalfig100}
\end{figure}

\begin{figure}
\includegraphics[width=\columnwidth]{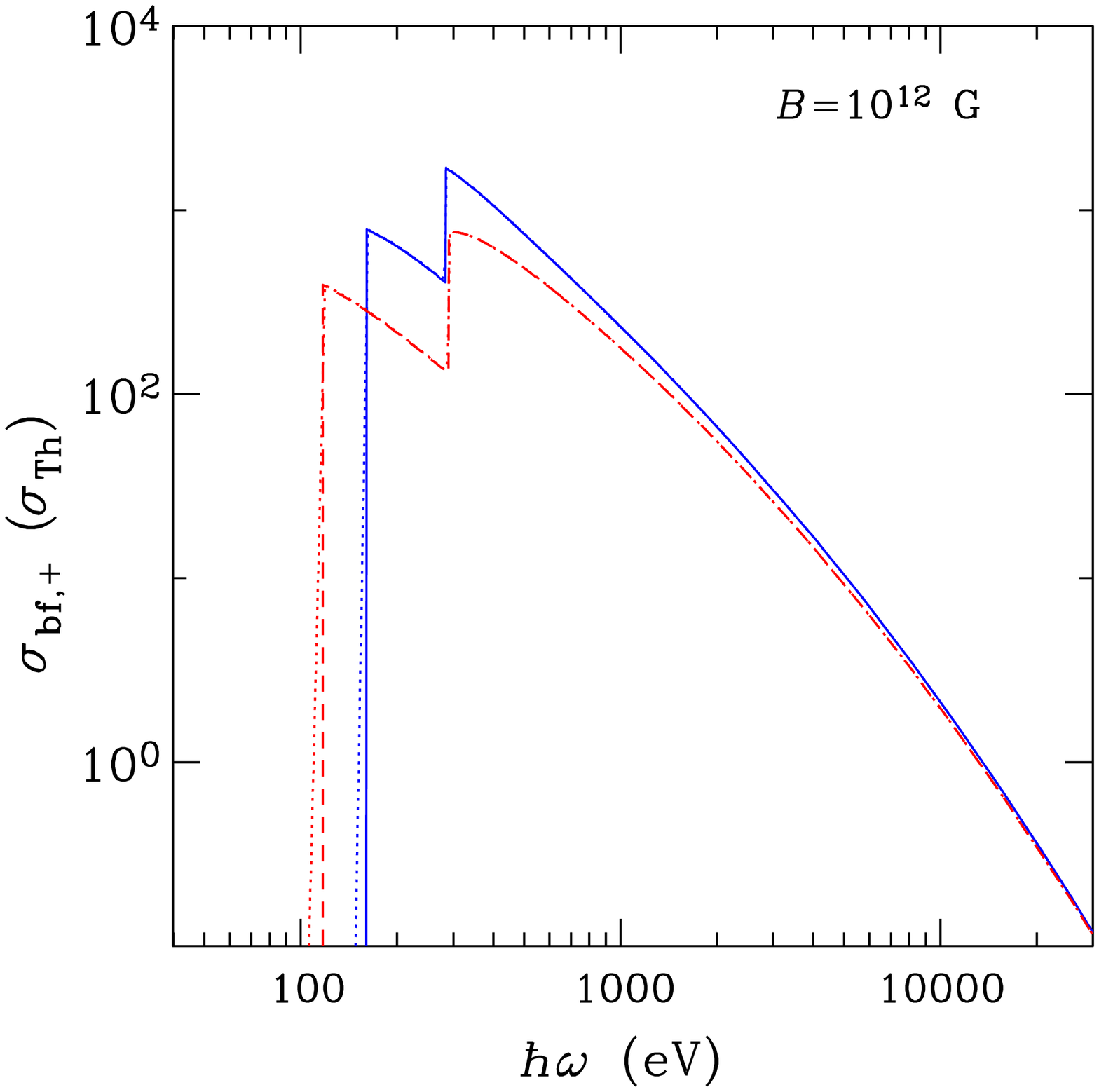}
\caption{Total cross section $\sigma_+$ versus photon energy for helium
photoionization, from initial states $(m_1,m_2)=(1,0)$
(solid lines)
 and $(2,0)$
(dashed lines). The
field strength is $10^{12}$ G.
The dotted
lines extending from each cross section curve represent the
effect of magnetic broadening on these cross sections, as approximated
in Eq.~(\ref{magbroad}),
for $T=10^6$~K.}
\label{sig12p}
\end{figure}

\begin{figure}
\includegraphics[width=\columnwidth]{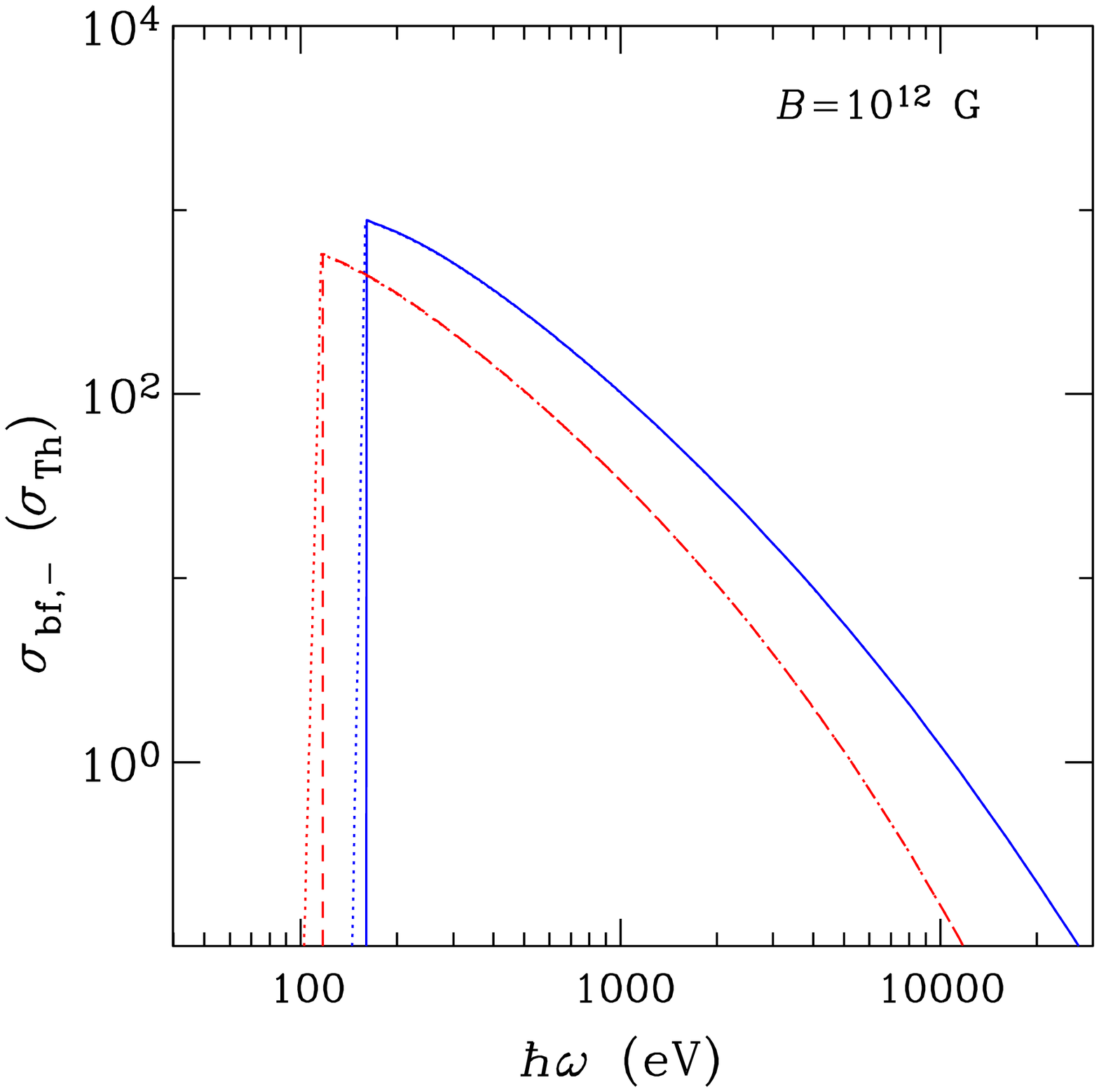}
\caption{The same as in Fig.~\ref{sig12p}, but for $\sigma_-$.}
\label{sig12m}
\end{figure}

\begin{figure}
\includegraphics[width=\columnwidth]{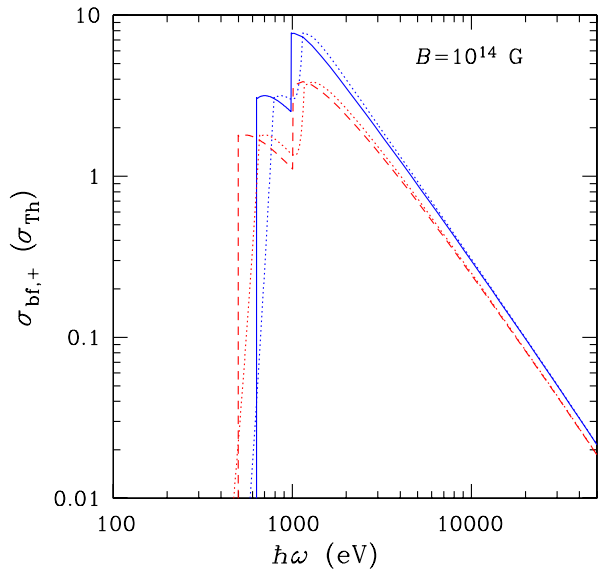}
\caption{The same as in Fig.~\ref{sig12p}, but for $B=10^{14}$~G.}
\label{sig14p}
\end{figure}

\begin{figure}
\includegraphics[width=\columnwidth]{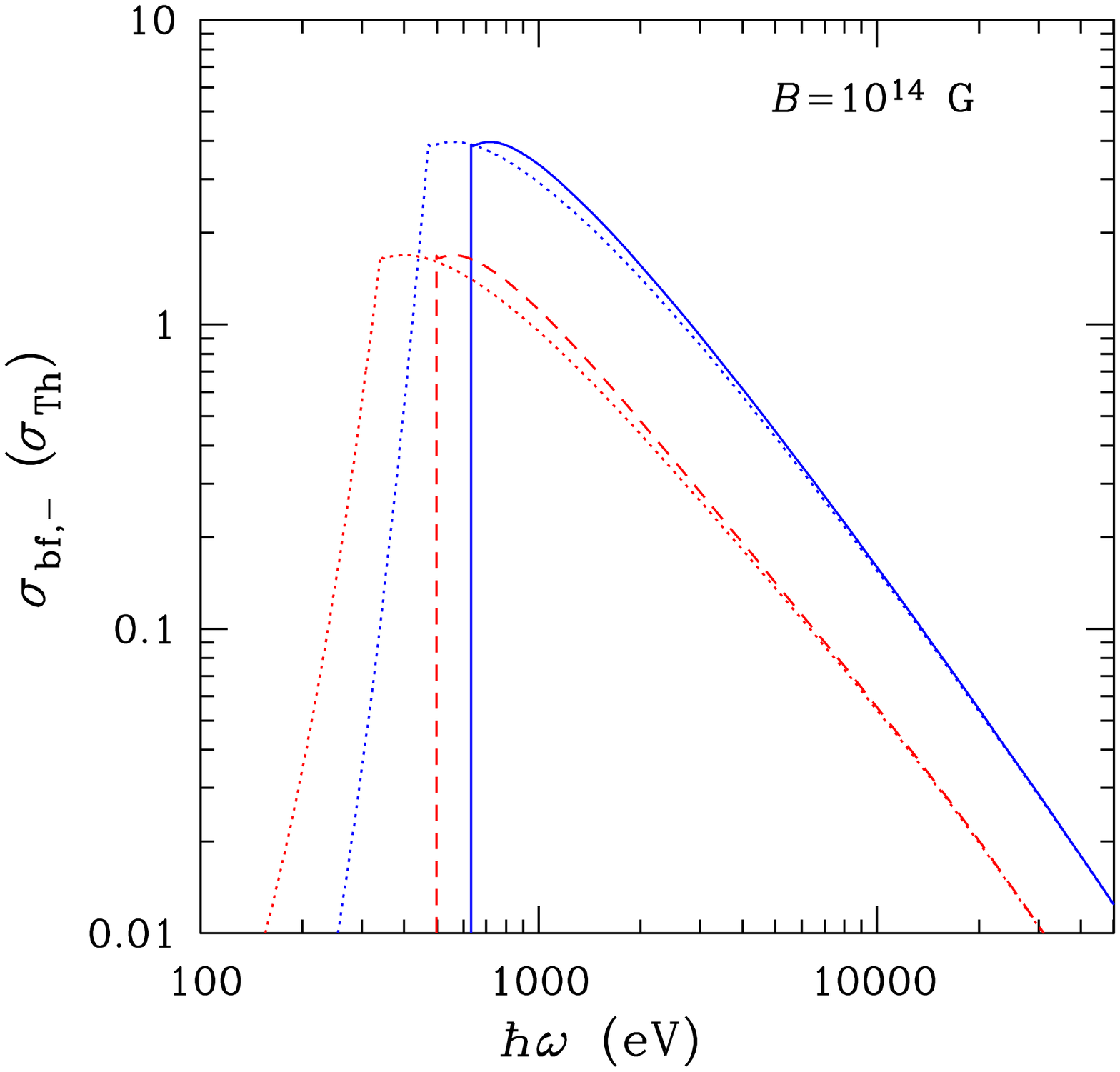}
\caption{The same as in Fig.~\ref{sig14p}, but for $\sigma_-$.}
\label{sig14m}
\end{figure}

Table~\ref{fittable} gives results (threshold photon energies 
and cross section fitting formulas, see below) for all possible 
bound-free transitions. 
Figure~\ref{partialfig} shows partial cross section curves for all 
bound-free transitions from the ground state of helium for $B_{12}=1$. 
The transition $\ket{00,10}\rightarrow\ket{0k,20}$
is an example of a `weak' transition, whose oscillator strength
is small because of the approximate orthogonality of one-electron
wave functions, as discussed at the end of Sect.\,\ref{sect:bb}.
It is included in this figure to confirm the accuracy of our assumption. 
Figures~\ref{totalfig} and \ref{totalfig100} show total cross section
curves for a photon polarized along the magnetic field, for $B_{12}=1$
and $100$ respectively.
Figures~\ref{sig12p} and \ref{sig12m} show total cross sections
for the circular polarizations, $\alpha=\pm$, for $B_{12}=1$.
Finally, Figs.~\ref{sig14p} and \ref{sig14m} show total cross sections
for $\alpha=\pm$ and $B_{12}=100$.

\subsection{Fitting Formula}

The high-energy cross section scaling relations from \citet{PP93}, 
which were derived for hydrogen photoionization
in strong magnetic fields, also hold for helium:
\be
\sigma_\mathrm{bf,0} \propto \left(\frac{1}{\hbar\omega}\right)^{2m_i+9/2}
\ee
\be
\sigma_{\mathrm{bf},\pm} \propto \left(\frac{1}{\hbar\omega}\right)^{2m_i+7/2} \,,
\ee
where $m_i$ is the $m$ value of the initial electron that transitions to the free state. In addition, we use similar fitting formulae for our numerical cross sections:
\be
\sigma_\mathrm{bf,0} \simeq \frac{{\cal C}}{(1+{\cal A}y)^{2.5} (1+{\cal B}(\sqrt{1+y}-1))^{4(m_i+1)}} 
\, \sigma_\mathrm{Th} \nonumber
\ee
\be
\sigma_{\mathrm{bf},\pm} \simeq \frac{{\cal C}(1+y)}{(1+{\cal A}y)^{2.5}
 (1+{\cal B}(\sqrt{1+y}-1))^{4(m_i+1)}} \, \sigma_\mathrm{Th}
\label{fitforms}
\ee
where $y = \varepsilon_f/\hbar\omega_{\rm thr}$ and
$\hbar\omega_{\rm thr}$ is the threshold photon energy for
photoionization. These formulas have been fit to the cross
section curves with respect to the free electron energy
$\varepsilon_f$ in approximately the $1$\,--\,$10^4$~eV range (the
curves are fit up to $10^5$~eV for strong magnetic fields
$B_{12}=50-100$, in order to obtain the appropriate high-energy
factor). The data points to be fit are weighted proportional to
their cross section values plus a slight weight toward low-energy
values, according to the formula (error in $\sigma$) $\propto
\sigma\,{\varepsilon_f}^{0.25}$.

Results for the three fitting parameters, ${\cal A}$, ${\cal B}$,
and ${\cal C}$, are given in Table~\ref{fittable} for various
partial cross sections over a range of magnetic field strengths.
For photoionization in strong magnetic fields ($B_{12}\ga 50$) the
cross section curves we generate for the $\sigma_+$ and $\sigma_-$
transitions have a slight deficiency at low electron energies, such
that the curves peak at $\varepsilon_f \simeq 10$~eV, rather than at
threshold as expected. These peaks do not represent a real effect, but
rather reflect the limits on the accuracy of our code (the overlap of
the wave function of the transitioning electron pre- and
post-ionization is extremely small under these conditions). Because
the cross section values are not correct at low energies, our fits are
not as accurate for these curves. In Table~\ref{fittable} we have
marked with a `$\ast$' those transitions which are most inaccurately
fit by our fitting formula, determined by cross section curves with
low-energy dips greater than $5\%$ of the threshold cross section
value.

\section{Finite nucleus mass effects}
\label{sect:motion}

So far we have used the infinite ion mass approximation. 
In this section we shall
evaluate the validity range of this approximation and
suggest possible corrections.

It is convenient to use the coordinate system which contains
the centre-of-mass coordinate $\bm{R}_\mathrm{cm}$
and the relative coordinates $\{\bm{r}_j\}$ of the electrons
with respect to the nucleus. 
Using a suitable canonical transformation,
the Hamiltonian $H$ of an arbitrary atom or ion can be separated
into three terms \citep{vb88,bv90,SchmCeder}: $H_1$ 
which describes
the motion of a free pseudo-particle with net charge $Q$
and total mass $M$ of the ion (atom), the coupling term $H_2$
between the collective and internal motion, and $H_3$ 
which describes the internal
relative motion of the electrons and the nucleus.
$H_1$ and $H_2$ are proportional to $M^{-1}$,
so they vanish in the infinite mass approximation.
It is important to note, however, that $H_3$ 
(the only non-zero term in the infinite mass
approximation) also contains a term that depends on $M_0^{-1}$,
where $M_0\approx M$ is the mass of the nucleus.
Thus, there are two kinds of non-trivial finite-mass effects:
the effects due to $H_1+H_2$,
which can be interpreted as caused by the electric field 
induced in the co-moving reference frame,
and the effects due to $H_3$, which arise
irrespective of the atomic motion.
Both kinds of effects have been 
included in calculations only for the H atom
\citep[][and references therein]{P94,PP97}
and He$^+$ ion \citep{BPV,PB05}.
For the He atom, only
the second kind of effects have been studied \citep{hujaj1,hujaj2}.

\subsection{Non-moving helium atom}
\label{sect:correction0}

The state of motion of an atom can be described
by pseudomomentum $\bm{K}$, which is a conserved vector
since $Q=0$ \citep[e.g.,][]{vb88,SchmCeder}. 
Let us consider first the non-moving helium atom: $K=0$.

According to \citet{hujaj1}, there are trivial
\emph{normal mass corrections}, which consist in the appearance of
reduced masses $m_e/(1\pm m_e/M_0)$ in $H_3$,
and non-trivial \emph{specific mass corrections},
which originate from the mass polarization operator.

The normal mass corrections for the total energy $E$
of the He state $\ket{m_1\nu_1,m_2\nu_2}$
can be described as follows:
\be
   E(M_0,B) =  \frac{E(\infty,(1+m_e/M_0)^2 B)}{1+m_e/M_0}
    + \hbar\Omc\sum_{j} m_j,
\label{E(M_0)}
\ee
where $\Omc=(m_e/M_0)\omc$
(for He, $\hbar\Omc=1.588 B_{12}$~eV).
The first term on the right-hand side describes 
the reduced mass transformation. The second term represents
the energy shift due to conservation of the total $z$ component
of the angular momentum. 
Because of this shift, the states with sufficiently large values
of $m_1+m_2$ become unbound (autoionizing, in analogy with the 
case of the H atom considered by \citealt{PPV}).
This shift is also important for radiative transitions
which change $(m_1+m_2)$ by $\Delta m\neq0$: the transition energy
$\hbar\omega_{ba}$ is changed by $\hbar\Omc\Delta m$.
The dipole matrix elements $M_{ba}$ are only slightly affected
by the normal mass corrections,
but 
the oscillator strengths are changed with changing $\omega_{ba}$
according to \req{oscstreq}.
The energy shift also
leads to the splitting of the photoionization threshold
by the same quantity $\hbar\Omc\Delta m$, with $\Delta m =0,\pm1$
depending on the polarization (in the dipole approximation).
Clearly, these corrections must be taken into account, 
unless $\Omc\ll\omega_{ba}$ or $\Delta m=0$, as illustrated
in the last two columns of Table~\ref{osctable}.

The specific mass corrections are more difficult to evaluate,
but they can be neglected in the considered $B$ range.
Indeed, calculations by \citet{hujaj1} show that these corrections
do not exceed 0.003 eV at $B\leq10^4B_0$.

\subsection{Moving helium atom}

Eigenenergies and wave functions of a moving atom depend
on its pseudomomentum $\bm{K}$
perpendicular to the magnetic field. This dependence
can be described by Hamiltonian components 
\citep[e.g.,][]{SchmCeder}
\be
   H_1+H_2 = \frac{K^2}{2M}
    + \sum_j \frac{e}{Mc} \bm{K}\cdot(\bm{B}\times\bm{r}_j),
\ee
where $\sum_j$ is the sum over all electrons.
The dependence on $K_z$ is trivial, but the dependence
on the perpendicular component $\bm{K}_\perp$ is not.
The energies depend on the absolute value $K_\perp$.
For calculation
of radiative transitions, it is important
to take into account that
the pseudomomentum of the atom in the initial and final state
differ due to recoil:
$\bm{K}'=\bm{K}+\hbar\bm{q}$. 
Effectively the recoil adds a term $\propto\bm{q}$ into 
the interaction operator \citep[cf.][]{PPV,PP97}.
The recoil should be
neglected in the dipole approximation.

The atomic energy $E$ depends on $K_\perp$ differently for
different quantum states of the atom. In a real neutron star
atmosphere, one should integrate the binding energies
and cross sections over the $K_\perp$-distribution
of the atoms, in order to obtain
the opacities.\footnote{For the hydrogen atom, this has been done
by \citet{PP95} for bound-bound transitions
and by \citet{PP97} for bound-free transitions.}
Such integration leads to the specific magnetic broadening
of spectral lines and ionization edges.
Under the conditions typical for neutron star atmospheres,
the magnetic broadening turns out to be much larger
than the conventional Doppler and collisional
broadenings \citep{PP95}.

At present the binding energies and cross sections
of a moving helium atom have not been calculated.
However, we can approximately estimate the magnetic broadening
for  $T \ll |(\Delta E)_\mathrm{min}|/\kB$, where
$(\Delta E)_\mathrm{min}$ is the energy difference
from a considered atomic level to the nearest level
admixed by the perturbation due to atomic motion, 
and
$\kB$ is the Boltzmann constant.
In this case, the $K_\perp$-dependence of $E$
can be approximated by the formula
\be
   E(K_\perp) = E(0) + \frac{K_\perp^2}{2M_\perp},
\ee
where $E(0)$ is the energy in the infinite mass approximation and
$M_\perp = K_\perp (\partial E/\partial K_\perp)^{-1}$ 
is an effective `transverse' mass,
whose value ($M_\perp > M$) depends on the quantum state considered
\citep[e.g.,][]{vb88,PM93}.

Generally, at every value of $K_\perp$
one has a different cross section $\sigma(\omega,K_\perp)$.
Assuming the equilibrium (Maxwell--Boltzmann) distribution
of atomic velocities, the $K_\perp$-averaged cross section
can be written as
\be
\sigma(\omega) = \int_{E_\mathrm{min}}^\infty
 \!\!\!\exp\left(\frac{E(0)-E(K_\perp)}{\kB T}\right) \sigma(\omega,K_\perp) \, 
\frac{\dd E(K_\perp)}{\kB T},
\label{sigma-int}
\ee
where $E_\mathrm{min}=-\hbar\omega$.

The transitions that were dipole-forbidden for an atom at rest
due to the conservation of the total $z$-projection of angular momentum
become allowed for a moving atom. Therefore, the selection rule
$\Delta m=\alpha$ [Eqs.~(\ref{seleq1})--(\ref{seleq3})]
does not strictly hold, and we must write
\be
   \sigma(\omega,K_\perp) = \sum_{m'}\sigma_{m'}(\omega,K_\perp),
\label{sigma-sum}
\ee
where the sum of partial cross sections is over all final quantum
numbers $m'$ (with $m'\geq 0$ and $m'\neq m_2$ for $\Delta\nu=0$)
which are energetically allowed. For bound-bound transitions,
this results in the splitting of an absorption line at a
frequency $\omega_{ba}$ in a multiplet at frequencies
$\omega_{ba}+\delta m \Omc +
(M_{\perp,m'}^{-1}-M_{\perp}^{-1})K_\perp^2/ 2\hbar$, where
$\delta m \equiv m'-m-\alpha$ and $M_{\perp,m'}$ is the
transverse mass of final states. For photoionization, we have
the analogous splitting of the threshold. In particular, there
appear bound-free transitions at frequencies $\omega <
\omega_\mathrm{thr}$ -- they correspond to $\delta m <
K_\perp^2/(2M_\perp\hbar\Omc)$. Here, $\omega_\mathrm{thr}$ is
the threshold in the infinite ion mass approximation, and one
should keep in mind that the considered perturbation theory is
valid for $K_\perp^2/2M_\perp \ll |(\Delta E)_\mathrm{min}| <
\hbar\omega_\mathrm{thr}$. According to \req{sigma-sum},
$\sigma(\omega,K_\perp)$ is notched at $\omega <
\omega_\mathrm{thr}$, with the cogs at partial thresholds
$\omega_\mathrm{thr}+\delta m \Omc - K_\perp^2/(2M_\perp\hbar)$
(cf.\ Fig.~2 in \citealt{PP97}).

Let us approximately evaluate the resulting envelope of the 
notched photoionization cross section (\ref{sigma-sum}),
assuming that the `longitudinal' matrix elements
[$\langle\ldots\rangle$ constructions in
Eqs.~(\ref{seleq1})--(\ref{seleq3})] do not depend on $K_\perp$.
The `transverse' matrix elements can be evaluated
following \citet{PP97}: in the perturbation approximation,
they are proportional to $|\xi|^{|\delta
m|}\mathrm{e}^{-|\xi|^2/2}$, where
$|\xi|^2=K_\perp^2\rho_0^2/(2\hbar^2)$.
Then
\ba&&\hspace*{-2em}
   \sigma(\omega<\omega_\mathrm{thr},K_\perp)
   \approx\sigma(\omega_\mathrm{thr},0)
    \exp\left[-\frac{M_\perp}{M}
    \frac{\omega_\mathrm{thr}-\omega}{\Omc}\right]
\nonumber\\&&\qquad\qquad\times
    \theta\left(\frac{K_\perp^2}{2M_\perp}
           -\hbar(\omega_\mathrm{thr}-\omega)\right),
\ea
where $\theta(x)$ is the step function.
A comparison of this approximation with numerical calculations
for the hydrogen atom \citep{PP97} shows that it gives the 
correct qualitative behaviour of $\sigma(\omega,K_\perp)$.
For a quantitative agreement, one should 
multiply the exponential argument by a numerical factor
$\sim 0.5$--2, depending on the state and polarization.
This numerical correction is likely due to the neglected
$K_\perp$-dependence of the longitudinal matrix elements.
We assume that this approximation can be used
also for the helium atom.
Using \req{sigma-int}, we obtain
\be
   \sigma(\omega) 
   \approx\sigma(\omega_\mathrm{thr})
   \exp\left[ - \frac{M_\perp}{M}
   \frac{\omega_\mathrm{thr}-\omega}{\Omc} -
   \frac{\hbar(\omega_\mathrm{thr}-\omega)}{\kB T} \right]
\label{magbroad}
\ee
for $\omega <\omega_\mathrm{thr}$.
Here the transverse mass $M_\perp$ can be evaluated 
by treating the coupling Hamiltonian $H_2$ as a perturbation,
as was done by \citet{PM93} for the H atom.
Following this approach,
retaining only the main perturbation terms 
according to the approximate orthogonality relation (\ref{ortho})
 and neglecting the difference
between $M$ and $M_0$, we obtain an estimate
\be
   \frac{M}{M_\perp} \approx 1 - \sum_{\alpha=\pm} \frac{\alpha}{2}
   \sum_{b(\Delta m=\alpha)} \frac{\omc f^\alpha_{ba}/(2\omega_{ba})}{
   1+\omega_{ba}/\Omc},
\label{Mperp}
\ee
where $\ket{a}$ is the considered bound state 
($|00,10\rangle$ or $|00,20\rangle$ for the examples
 in Figs.~\ref{totalfig}--\ref{sig14m})
and $\ket{b}$
are the final bound states to which $\alpha=\pm$ transitions
$\ket{a}\to\ket{b}$ are allowed.
According to \req{f_approx},
the numerator in \req{Mperp} is close to $m+1$ for $\alpha=+$
and to $m$ for $\alpha=-$.

For the transitions from the ground state with polarization
$\alpha=-$, which are strictly forbidden in the infinite ion
mass approximation, using the same approximations as above we
obtain the estimate
$\sigma_-(\omega)\propto\sigma_+(\omega)\hbar\Omc\kB T
/ (\kB T + \hbar\Omc)^2$.

Examples of the photoionization envelope approximation, as
described in Eq.~(\ref{magbroad}) above, are shown in
Figs.~\ref{totalfig}--\ref{sig14m}.
In Figs.~\ref{sig14p} and \ref{sig14m} (for $B=10^{14}$~G), in
addition to the magnetic broadening, we see a significant shift
of the maximum, which originates from the last term in
\req{E(M_0)}. Such shift is negligible in  Figs.~\ref{sig12p} and
\ref{sig12m} because of the relatively small $\Omc$ value for
$B=10^{12}$~G.

Finally, let us note that the Doppler and collisional broadening
of spectral features in a strong magnetic field can be estimated,
following \citet{PM93}, \citet{PP95} and \citet{RRM}.
The Doppler spectral broadening profile is
\be
\phi_\mathrm{D}(\omega) = \frac{1}{\sqrt{\pi}\Delta
\omega_\mathrm{D}} 
\exp\left[-\frac{(\omega-\omega_0)^2}{\Delta
\omega_\mathrm{D}^2}\right],
\ee
with
\be
\Delta \omega_D = \frac{\omega_0}{c}\sqrt{\frac{2T}{M}}
 \left[\cos^2 \theta_B
  + \frac{M_\perp}{M} \sin^2 \theta_B\right]^{-1/2} ,
\ee
where $\theta_B$ is the angle between the wave vector and
$\bm{B}$.
The collisional broadening is given by
\be
\phi_\mathrm{coll}(\omega) = \frac{\Lambda_\mathrm{coll}}{2\pi}\frac{1}{(\omega-\omega_0)^2+(\Lambda_\mathrm{coll}/2)^2} \,,
\ee
with
\ba&&\hspace*{-2em}
\hbar\Lambda_\mathrm{coll} = 4.8 n_e a_0 r_{\rm eff}^2
\left(\frac{\kB T}{\rm Ryd}\right)^{1/6}
\nonumber\\&&
 = 41.5
\frac{n_e}{\mbox{$10^{24}$ cm$^{-3}$}} T_6^{1/6}
\left(\frac{r_{\rm eff}}{a_0}\right)^2 \mbox{ eV} ,
\ea
where $n_e$ is the electron number density and $r_{\rm eff}$ is
an effective electron-atom interaction radius, which is about 
the quantum-mechanical size of the atom. The convolution of the
Doppler, collisional and magnetic broadening profiles gives the
total shape of the cross section.
For bound-free transitions, the Doppler and collisional factors
can be neglected, but for the bound-bound transitions
they give the correct blue wings of the spectral features
\citep[cf.][]{PP95}.

\section{Conclusion}

We have presented detailed numerical results and fitting formulae
for the dominant radiative transitions (both bound-bound and
bound-free) of He atoms in strong magnetic fields in the range
of $10^{12}-10^{14}$~G. These field strengths may be most appropriate
for the identification of spectral lines observed 
in thermally emitting isolated neutron stars (see Sect.\,\ref{sec:intro}).

While most of our calculations are based on the 
infinite-nucleus-mass approximation, we have examined the effects 
of finite nucleus mass and atomic motion on the opacities. We found that for
the field strengths considered in this paper ($B\la 10^{14}$~G), these
effects can be incorporated into the infinite-mass results 
to obtain acceptable He opacities for neutron star atmosphere modelling.
For large field strengths, more accurate calculations of the energy levels
and radiative transitions of a moving He atom will be needed
in order to obtain reliable opacities.

\section*{Acknowledgments}
This work has been supported in part by NSF grant AST 0307252
and \textit{Chandra} grant TM6-7004X (Smithsonian Astrophysical Observatory). 
The work of A.P.\ is supported in part by FASI (Rosnauka)
grant NSh-9879.2006.2
and RFBR grants 05-02-16245 and 05-02-22003.

\end{document}